\documentclass[journal=jctcce,manuscript=article]{achemso}

\usepackage{graphicx}
\usepackage{dcolumn}
\usepackage{bm}
\usepackage{lipsum}
\usepackage[table,xcdraw]{xcolor}
\usepackage{array}
\usepackage{multirow}
\usepackage{soul}
\usepackage{chemformula}
\usepackage{xr}
\usepackage{hhline}
\usepackage{braket}
\usepackage{listings}
\usepackage{xspace}
\usepackage{threeparttable}
\usepackage{amsmath}
\usepackage{amssymb}

\usepackage[noabbrev]{cleveref}
\crefname{equation}{eq}{eqs}
\creflabelformat{equation}{#2#1#3}

\usepackage{algorithm}
\usepackage{algpseudocode}

\algrenewcommand\algorithmicrequire{\textbf{Input:}}
\algrenewcommand\algorithmicensure{\textbf{Output:}}
\algrenewcommand\algorithmiccomment[1]{\hfill{\color{gray}\(\triangleright\) #1}}

\newcommand{\bvec}[1]{\bm{\mathrm{#1}}}
\newcommand{\adj}{^{\dagger}}

\newcommand{\lgc}{^{\mathrm{L}}}

\newcommand{\plc}{^{\mathrm{P}}}
\newcommand{\hartree}{E_{\mathrm{h}}}
\newcommand{\sunit}{$E_{\mathrm{h}}^{-2}$}
\newcommand{\wn}{cm$^{-1}$}
\newcolumntype{B}{>{\bfseries}l}

\definecolor{goodorange}{RGB}{225,125,0}
\definecolor{goodgreen}{RGB}{0,125,0}
\definecolor{goodred}{RGB}{220,50,25}
\definecolor{goodblue}{RGB}{25,25,150}

\definecolor{codegreen}{rgb}{0,0.6,0}
\definecolor{codegray}{rgb}{0.5,0.5,0.5}
\definecolor{codepurple}{rgb}{0.58,0,0.82}
\definecolor{backcolour}{rgb}{0.95,0.95,0.92}

\lstdefinestyle{mystyle}{
    backgroundcolor=\color{backcolour},   
    commentstyle=\color{codegreen},
    keywordstyle=\color{magenta},
    numberstyle=\tiny\color{codegray},
    stringstyle=\color{codepurple},
    basicstyle=\ttfamily\footnotesize,
    breakatwhitespace=false,         
    breaklines=true,                 
    captionpos=b,                    
    keepspaces=true,                 
    numbers=left,                    
    numbersep=5pt,                  
    showspaces=false,                
    showstringspaces=false,
    showtabs=false,                  
    tabsize=2
}

\title{One-Step Relativistic Driven Similarity Renormalization Group Multireference Perturbation Theory}

\author{Zijun Zhao}
\email{zijun.zhao@emory.edu}
\author{Francesco A. Evangelista}
\email{francesco.evangelista@emory.edu}
\affiliation{Department of Chemistry and Cherry Emerson Center for Scientific Computation, Emory University, Atlanta, Georgia, 30322, United States}

\begin{document}

\begin{tocentry}
\includegraphics[width=3.25in]{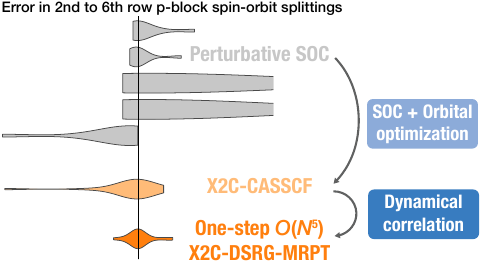}
\end{tocentry}

\begin{abstract}
We present an efficient implementation of a one-step relativistic second-order multireference perturbation theory based on the multireference driven similarity renormalization group (MR-DSRG) using the exact two-component (X2C) Hamiltonian, which we denote X2C-DSRG-MRPT2.
We show that the X2C-DSRG-MRPT2 method can accurately capture spin--orbit coupling (SOC) effects in the electronic structure of strongly correlated systems containing elements across the periodic table.
We further demonstrate that the X2C-DSRG-MRPT2 method, through its variational treatment of SOC effects, can yield spin--orbit splittings with mean absolute percentage errors consistently below 7\% with respect to experimental values for systems containing up to sixth row elements.
With its modest computational scaling (fifth power in system size) and high accuracy, X2C-DSRG-MRPT2 provides a promising avenue for the routine treatment of relativistic effects in strongly correlated molecular systems.
\end{abstract}

\maketitle
\section{Introduction}
The simultaneous treatment of both relativistic effects and strong electron correlation that is both economical and reliable is essential for scalable and accurate simulations of molecules containing heavy elements.
Relativistic effects--including scalar- and vector-relativistic effects\cite{hoyer.2024.10.1021/acs.jctc.4c00797}--are ubiquitous in heavy-element chemistry, playing a crucial role in spectroscopic properties, bonding characteristics, and reactivity patterns.\cite{pyykko.1979.10.1021/ar50140a002,pyykko.1988.10.1021/cr00085a006,pyykko.2012.10.1021/cr200042e,pyykko.2012.10.1146/annurev-physchem-032511-143755}
The past decades saw great research interest and ensuing advancements in relativistic quantum chemistry, including the algorithmic refinement of four-component (4C) methods based on the Dirac equation, enabling its routine use in high-accuracy computations;
\cite{abe.2006.10.1063/1.2404666,banerjee.2023.10.1063/5.0161871,bates.2015.10.1063/1.4906344,fleig.2003.10.1063/1.1590636,fleig.2006.10.1063/1.2176609,hoyer.2023.10.1063/5.0133741,jensen.1996.10.1063/1.471644,kelley.2013.10.1063/1.4807612,knecht.2010.10.1063/1.3276157,reynolds.2018.10.1063/1.5036594,saue.1999.10.1063/1.479958,shiozaki.2015.10.1021/acs.jctc.5b00754,sun.2021.10.1021/acs.jctc.1c00137,ten-no.2012.10.1063/1.4757415,thyssen.2008.10.1063/1.2943670,visscher.1995.10.1002/qua.560560844,visscher.1996.10.1063/1.472655,visscher.1997.10.1007/s002140050280,visscher.2001.10.1063/1.1415746,watanabe.2002.10.1063/1.1476694} 
the development and commonplace adoption of the exact two-component (X2C) Hamiltonian that can accurately capture one-electron relativistic effects at a fraction of the cost of the 4C methods,
\cite{ilias.2007.10.1063/1.2436882,kutzelnigg.2006.10.1080/00268970600662481,li.2012.10.1063/1.4758987,li.2014.10.1063/1.4891567,liu.2006.10.1063/1.2222365,liu.2007.10.1063/1.2710258,liu.2009.10.1063/1.3159445,peng.2007.10.1063/1.2772856} 
as well as further developments for the inclusion of two-electron relativistic effects;
\cite{peng.2007.10.1063/1.2772856,boettger.2000.10.1103/PhysRevB.62.7809,ehrman.2023.10.1021/acs.jctc.3c00479,knecht.2022.10.1063/5.0095112,liu.2018.10.1063/1.5023750,peng.2012.10.1007/s00214-011-1081-y,sikkema.2009.10.1063/1.3239505,surjuse.2026.10.1021/acs.jctc.6c00032,zhang.2022.10.1021/acs.jpca.2c02181} 
and, more recently, the resurgence of the state-interaction spin--orbit (SISO) approach that adds spin--orbit coupling (SOC) effects as a perturbation to spin-free wave functions, enabling even more scalable relativistic computations.\cite{hoyer.2024.10.1021/acs.jctc.4c00797,jangid.2025.10.1021/acs.jctc.5c01633,liao.2024.10.1021/acs.jpca.3c08031,majumder.2024.10.1021/acs.jctc.4c00458,park.2025.10.1021/acs.jctc.5c01543,wang.2025.10.3390/molecules30092082,zhai.2022.10.1063/5.0107805,fedorov.2000.10.1063/1.481136,berning.2000.10.1080/00268970009483386,malmqvist.2002.10.1016/S0009-26140200498-0,sayfutyarova.2016.10.1063/1.4953445}

The treatment of relativistic effects can be grouped into two categories: one-step and two-step approaches.
In the one-step approaches, relativistic effects are incorporated either variationally or perturbatively \emph{throughout the computation}.
This typically allows for a more balanced treatment of relativity and electron correlation, for example, by optimizing the molecular orbitals and the correlated wave function in the presence of SOC, and has been shown to be superior to two-step approaches in cases of a dense manifold of low-lying states strongly coupled by SOC.\cite{zhai.2022.10.1063/5.0107805}
In the two-step approaches, on the other hand, relativistic effects are added later in the computation, typically after orbital optimization, so that real-valued orbitals can be used.
The SOC Hamiltonian could be introduced either at the electron correlation step, or even after the correlation treatment has been performed, where a set of spin-free states are mixed into spin--orbit coupled states through diagonalizing either a phenomenological or \emph{ab initio} SOC Hamiltonian in the basis of the spin-free states.
The two-step approaches have the advantage of being more computationally affordable, firstly through the use of real arithmetic and spatial orbitals in parts of the computation, secondly owing to the fact that the SOC treatment is typically limited to a small number of low-lying states of interest.

The aforementioned advances have also been frequently paired with multireference electron correlation methods, owing to the prevalence of quasi-degeneracies in heavy-element chemistry, necessitating the use of a multi-determinantal reference wave function.
These include four-component internally-contracted perturbation theories and multireference configuration interaction (MRCI);\cite{shiozaki.2015.10.1021/acs.jctc.5b00754} and state-interaction approaches based on complete active space self-consistent field (CASSCF-SO),\cite{fedorov.2000.10.1063/1.481136,berning.2000.10.1080/00268970009483386,malmqvist.2002.10.1016/S0009-26140200498-0} CAS second-order perturbation theory (CASPT2-SO),\cite{malmqvist.1989.10.1016/0009-26148985347-3} density matrix renormalization group (DMRG),\cite{sayfutyarova.2016.10.1063/1.4953445} $N$-electron valence second-order perturbation theory (NEVPT2),\cite{majumder.2024.10.1021/acs.jctc.4c00458} and linearized pair-density functional theory (L-PDFT).\cite{jangid.2025.10.1021/acs.jctc.5c01633}

There is, however, a distinct scarcity of one-step two-component multireference methods.
Notable examples include the semi-stochastic heat-bath configuration interaction (SHCI) implementation by Mussard and Sharma;\cite{mussard.2018.10.1021/acs.jctc.7b01019} the uncontracted (uc) X2C-MRCI\cite{hu.2020.10.1021/acs.jctc.9b01290} and second-order perturbation theory (uc-X2C-MRPT2)\cite{lu.2022.10.1021/acs.jctc.2c00171} by Li and co-workers; the multi-configurational PDFT (X2C-MC-PDFT) by the Li, Gagliardi, and Truhlar groups;\cite{sharma.2022.10.1021/acs.jctc.2c00062} and the X2C-DMRG by Li and co-workers.\cite{hoyer.2022.10.1021/acs.jpca.2c02150}
With the exception of the one-step relativistic SHCI method, the other methods all use the X2C-CASSCF method to obtain the reference wave function,\cite{jenkins.2019.10.1021/acs.jctc.9b00011a} whose recent introduction and implementation has opened the door to facile development of more one-step two-component multireference methods.

The multireference driven similarity renormalization group (MR-DSRG) formalism is a promising approach for the treatment of strong electron correlation, as it encompasses a family of systematically improvable methods, has low polynomial scaling, and is free from the intruder-state problem.\cite{li.2019.10.1146/annurev-physchem-042018-052416}
The second- and third-order perturbation theories based on the MR-DSRG framework (DSRG-MRPT2/3) have already been adapted to four-component Hamiltonians by the present authors,\cite{zhao.2024.10.1021/acs.jpclett.4c01372} and show great promise in accurately capturing SOC effects in heavy-atom-containing strongly correlated systems.
These theories require only up to three-body reduced density cumulants.
DSRG-MRPT2 has also additionally been paired with the state-interaction approach very recently,\cite{park.2025.10.1021/acs.jctc.5c01543,wang.2025.10.3390/molecules30092082} with results that compare favorably to other two-step methods based on NEVPT2, for example.

In this work, we present an efficient implementation of a one-step relativistic multireference perturbation theory based on the driven similarity renormalization group second-order multireference perturbation theory using the exact two-component Hamiltonian (X2C-DSRG-MRPT2).
The present formalism can deliver valence excited states \textit{via} the state-averaging formalism,\cite{li.2018.10.1063/1.5019793} and derives its efficiency from using the density fitting approximation.\cite{hannon.2016.10.1063/1.4951684}
It incorporates relativistic effects fully variationally at the multi-configurational self-consistent field (MCSCF) level, and, as we will show, it accurately captures SOC effects in strongly correlated systems containing elements across the periodic table, rivaling other state-of-the-art relativistic multireference theories.
Efficient implementations of the X2C-CASSCF and X2C-DSRG-MRPT2 methods are available in the latest version of \textsc{Forte2},\cite{forte2} a standalone open-source suite of quantum chemistry methods for strongly correlated molecular systems.

The rest of the paper is organized as follows.
In \cref{sec:theory}, we briefly review the components of the X2C-DSRG-MRPT2 method, including the X2C Hamiltonian and the X2C-CASSCF method, and then present a family of approximations to the X2C-DSRG-MRPT2 method that differ in the level of treatment of SOC effects.
In \cref{sec:comp_details}, we provide computational details for the computations presented in this work.
In \cref{sec:results}, we present the main results of this work, including the spin--orbit splittings of up to sixth-row $p$-block atoms in \cref{sec:pblock}, which also include a detailed analysis of the different approximate schemes of the X2C-DSRG-MRPT2 method using the $p$-block atoms as a test case; the spin--orbit splittings of a set of transition metal atoms in \cref{sec:tm}; the spin--orbit splittings of a set of open-shell diatomic molecules in \cref{sec:diatomics}; and the potential energy surfaces of the TlH molecule in \cref{sec:pes}.
Finally, we conclude in \cref{sec:conclusion} with a summary and outlook for future work.

\section{Theory}
\label{sec:theory}
\subsection{The One-Electron SNSO-X2C Hamiltonian}
The starting point of the exact two-component (X2C) formalism is the matrix modified one-electron Dirac equation, given by
\begin{equation}
    \label{eq:mod_dirac}
    \underbrace{
    \begin{bmatrix}
        \bvec{V} & \bvec{T} \\
        \bvec{T} & \bvec{W}/4c^2-\bvec{T}
    \end{bmatrix}}_{\bvec{h}^{\mathrm{D}}}
    \begin{bmatrix}
        \bvec{C}\lgc \\
        \bvec{C}\plc
    \end{bmatrix}=
    \underbrace{
    \begin{bmatrix}
        \bvec{S} & \bvec{0} \\
        \bvec{0} & \bvec{T}/2c^2
    \end{bmatrix}}_{\bvec{G}}
    \begin{bmatrix}
        \bvec{C}\lgc \\
        \bvec{C}\plc
    \end{bmatrix}E,
\end{equation}
where $\bvec{h}^{\mathrm{D}}$ is the modified Dirac Hamiltonian, $\bvec{G}$ is the metric matrix, and $\bvec{C}^{\mathrm{L}}$ ($\bvec{C}^{\mathrm{P}}$) is the large (pseudo-large) component coefficient matrix.
\Cref{eq:mod_dirac} has been expanded in a finite basis set, and implies that the restricted kinetic balance (RKB) condition is used to construct the small component basis functions.\cite{liu.2010.10.1080/00268971003781571,saue.2011.10.1002/cphc.201100682}
The matrices $\bvec{V}$, $\bvec{T}$, $\bvec{S}$ are the standard potential, kinetic, and overlap integrals, respectively, while $\bvec{W}$ is the modified potential matrix that includes the SOC effects, defined as
\begin{equation}
    W_{\mu\nu}=\braket{\chi_{\mu}|\bvec{p}\hat{V}\cdot\bvec{p}|\chi_{\nu}} +i\braket{\chi_{\mu}|\bvec{\sigma}\cdot(\bvec{p}\hat{V}\times\bvec{p})|\chi_{\nu}},
\end{equation}
where $\bvec{p}$ is the momentum operator, and $\bvec{\sigma}$ is the 3-vector of Pauli spin matrices.
The omission of the second term in the definition of $\bvec{W}$ results in the spin-free X2C (sf-X2C) Hamiltonian,\cite{cheng.2011.10.1063/1.3624397} which only captures scalar-relativistic effects, but allows for the interface with virtually any real-valued non-relativistic electronic structure method.
The X2C formalism seeks a transformation that block-diagonalizes the modified Dirac Hamiltonian, decoupling the positive-energy (electronic) solutions from the negative-energy (positronic) ones:
\begin{equation}
    \bvec{U}\adj\bvec{h}^{\mathrm{D}}\bvec{U}=
    \begin{bmatrix}
        \bvec{h}^+ & \bvec{0} \\
        \bvec{0} & \bvec{h}^-
    \end{bmatrix},
\end{equation}
where $\bvec{U}$ is the X2C decoupling transformation matrix, and $\bvec{h}^+$ and $\bvec{h}^-$ are the resulting two-component Hamiltonians for the positive- and negative-energy solutions, respectively.
The form that this transformation takes has been derived in detail by Ilia\v{s} and Saue \cite{ilias.2007.10.1063/1.2436882} and Liu and Peng,\cite{liu.2006.10.1063/1.2222365,liu.2007.10.1063/1.2710258,peng.2007.10.1063/1.2772856,liu.2009.10.1063/1.3159445} and we refer the reader to these works for details.

The X2C Hamiltonian $\bvec{h}^+$ includes one-electron relativistic effects, and is typically used in conjunction with the \emph{untransformed} non-relativistic two-electron integrals, making it possible to adapt it to almost any electronic structure method.
The resulting molecular electronic Hamiltonian is given by
\begin{equation}
    \hat{H} = \sum_{pq} h^{q,+}_p \hat{a}_p\adj \hat{a}_q + \frac{1}{4}\sum_{pqrs} v^{rs,\mathrm{nr}}_{pq} \hat{a}_p\adj \hat{a}_q\adj \hat{a}_s \hat{a}_r,
\end{equation}
where $p,q,r,s$ are two-spinor indices, $h^{q,+}_p$ are the matrix elements of the X2C Hamiltonian $\bvec{h}^+$, $v^{rs,\mathrm{nr}}_{pq}$ are the non-relativistic antisymmetrized two-electron integrals, and $\hat{a}_p\adj$ and $\hat{a}_p$ are the standard fermionic creation and annihilation operators, respectively.
The errors arising from neglecting to transform the two-electron term (to do so exactly would incur the same cost as four-component theories\cite{dyall.2007.10.1093/oso/9780195140866.001.0001,sikkema.2009.10.1063/1.3239505,saue.2011.10.1002/cphc.201100682}) is sometimes termed the two-electron picture change (2ePC) errors, which amount to the neglect of two-electron spin-spin and spin--orbit interactions.\cite{saue.2011.10.1002/cphc.201100682}
In this work, we account for the 2ePC errors by empirically scaling the spin-dependent part of the one-electron Hamiltonian $\bvec{h}^+$:
\begin{equation}
    \label{eq:snso}
    h^{+,\mathrm{SD}}_{\mu\nu}\leftarrow\left(1-\sqrt{\frac{Q(n_{\mu},l_{\mu})Q(n_{\nu},l_{\nu})}{Z_{\mu}Z_{\nu}}}\right)h^{+,\mathrm{SD}}_{\mu\nu},
\end{equation}
where $\bvec{h}^{+,\mathrm{SD}} =\bvec{h}^+-\frac{1}{2}(\bvec{h}^{+}_{\alpha\alpha}+\bvec{h}^{+}_{\beta\beta})\otimes\bvec{I}_2$ is the spin-dependent part of $\bvec{h}^{+}$, while $\bvec{h}^+_{\sigma\sigma'}$ ($\sigma\in\{\alpha,\beta\}$) are the spin blocks of the X2C Hamiltonian $\bvec{h}^+$; $Z_{\mu}$ is the nuclear charge of the atom that basis function $\chi_{\mu}$ is associated with, and $Q$ are empirical parameters for the screening of one-electron SOC by two-electron SOC, and are generally dependent on both the angular momentum, $l$, of the basis function, and the row-number, $n$, of the atom associated with the basis function.
This empirical scaling scheme, also known as the screened-nuclear spin--orbit (SNSO) approximation, was first proposed by Boettger, \cite{boettger.2000.10.1103/PhysRevB.62.7809} and re-parameterized by Ehrman \textit{et al.} \cite{ehrman.2023.10.1021/acs.jctc.3c00479} with benchmark results from four-component Dirac--Hartree--Fock theory using the Coulomb--Breit two-electron operator, which carries the full two-electron spin-spin, spin--orbit, and orbit-orbit couplings.
We summarize the algorithm used to construct the SNSO-X2C Hamiltonian $\bvec{h}^+$ in \cref{alg:x2c}, specifically, we follow the algorithm outlined in ref~\citenum{liu.2009.10.1063/1.3159445}.

Another popular way of accounting for 2ePC errors is through atomic mean-field approaches.\cite{ilias.2001.10.1063/1.1413510,knecht.2022.10.1063/5.0095112,zhang.2022.10.1021/acs.jpca.2c02181}
These have been shown to be accurate and cost-effective, and we will explore these approaches in the future.
However, they typically require evaluating atomic two-electron relativistic integrals and solving the atomic four-component Dirac--Hartree--Fock equations, which would add considerable complexity to the implementation.
Finally, the molecular mean-field approach developed by Sikkema \textit{et al.}\cite{sikkema.2009.10.1063/1.3239505} has also enjoyed considerable popularity due to its conceptual simplicity and good performance,\cite{zhang.2024.10.1021/acs.jpca.3c08167,yuwono.2025.10.1063/5.0248535} however, it requires the solution of the molecular four-component Dirac--Hartree--Fock equation as a prerequisite, which increases the overall computational demands.

To control the linear dependence of the basis set, especially when working with decontracted basis sets, we obtain an orthonormal basis by diagonalizing the overlap matrix $\bvec{S}$ and discarding eigenvectors corresponding to eigenvalues $s_i$ smaller than $\eta s_{\mathrm{max}}$, where $s_{\mathrm{max}}$ is the largest eigenvalue of $\bvec{S}$, and $\eta$ is a user-defined threshold, which is set to $10^{-8}$ in this work.
The same threshold is also consistently applied in downstream SCF computations.
To solve \cref{eq:mod_dirac}, the $\bvec{T}/2c^2$ matrix is then transformed to the orthonormal basis, and the resulting generalized eigenvalue problem is solved.
All subsequent steps are performed in the orthonormal basis with appropriately transformed quantities, and the final X2C Hamiltonian $\bvec{h}^+$ is transformed back to the original atomic basis, \emph{before} the SNSO scaling is applied.
This procedure ensures that any elimination of linear dependence in the basis set is carried out consistently across the large and pseudo-large components, and thus preserves the RKB condition.

\begin{algorithm}[H]
\caption{Algorithm used for the construction of the screened-nuclear spin--orbit X2C Hamiltonian.}
\label{alg:x2c}
\begin{algorithmic}[1]
\Require Basis set $\{\chi_{\mu}\}$, nuclear positions $\{\bvec{R}_I\}$, speed of light $c$
\Ensure X2C Hamiltonian $\bvec{h}^+$ (optionally in contracted basis)

\State $\bvec{P}, \{\xi_{\mu}\} \gets \textsc{DecontractBasis}(\{\chi_{\mu}\})$
\Comment{$\bvec{P}$ is the transformation matrix between contracted and decontracted bases: see line 13}

\State $\bvec{S}, \bvec{T}, \bvec{V}, \bvec{W} \gets \textsc{IntegralEngine}(\{\xi_{\mu}\},\{\bvec{R}_I\})$

\State $\bvec{h}^{\mathrm{D}}, \bvec{G} \gets \textsc{BuildDiracEq}(\bvec{S}, \bvec{T}, \bvec{V}, \bvec{W}, c)$

\State $\bvec{E},\bvec{C} \gets \textsc{eigh}(\bvec{h}^{\mathrm{D}}, \bvec{G})$
\Comment{Solve the modified Dirac equation}

\State $\bvec{X} \gets \bvec{C}_+^{\mathrm{P}}\left(\bvec{C}_+^{\mathrm{L}}\right)^{-1}$
\Comment{The coupling matrix between the pseudo-large and the large component}

\State $\widetilde{\bvec{S}} \gets \bvec{S} + \dfrac{1}{2c^2}\bvec{X}\adj \bvec{T}\bvec{X}$

\State $\bvec{\Lambda}, \bvec{Z} \gets \textsc{eigh}(\widetilde{\bvec{S}}, \bvec{S})$

\State $\bvec{R} \gets \bvec{Z}\bvec{\Lambda}^{-1/2}\bvec{Z}\adj \bvec{S}$
\Comment{The X2C picture-change matrix}

\State $\bvec{h}^{\mathrm{NESC}} \gets
\bvec{V}
+ \bvec{T}\bvec{X}
+ \bvec{X}\adj \bvec{T}
+ \bvec{X}\adj(\bvec{W}/4c^2-\bvec{T})\bvec{X}$

\State $\bvec{h}^+ \gets \bvec{R}\adj \bvec{h}^{\mathrm{NESC}} \bvec{R}$

\State $\bvec{h}^+ \gets \textsc{ApplySNSOScaling}(\bvec{h}^+)$
\Comment{Apply the SNSO scaling [\textit{via} Eq.~5]}

\If{recontract}
    \State $\bvec{h}^+ \gets \bvec{P}\adj\bvec{h}^+\bvec{P}$
    \Comment{Project into contracted basis}
\EndIf
\end{algorithmic}
\end{algorithm}

\subsection{X2C-CASSCF}
The X2C-CASSCF method recently introduced by Jenkins \textit{et al.}\cite{jenkins.2019.10.1021/acs.jctc.9b00011a} is the first of its kind to use the full (spin-free and spin-dependent) X2C Hamiltonian variationally in a multiconfigurational self-consistent field (MCSCF) computation.
Several other implementations of 2C-CASSCF have been reported in the literature since, that use different one-electron spin--orbit Hamiltonians,\cite{guo.2026.10.48550/arXiv.2603.03013,wang.2026.10.1021/acs.jctc.6c00321} and all of them can, in principle, be used to optimize two-component CASSCF wave functions with arbitrary one-electron spin--orbit Hamiltonians.
The formulation of X2C-CASSCF closely follows that of the non-relativistic CASSCF method.
The complete active space (CAS) wave function is given by:
\begin{equation}
    \label{eq:cas_ref}
    \ket{\Psi_0} = \sum_{\mu=1}^{d} c_{\mu} \ket{\Phi_{\mu}},
\end{equation}
where $c_{\mu}$ are the complex CI coefficients, and $\ket{\Phi_{\mu}}$ are Slater determinants constructed from a set of two-component molecular spinors $\{\phi_p\}$, given by:
\begin{equation}
    \phi_p = \begin{pmatrix}
        \sum_{\mu}C^{\mu,\alpha}_{p}\chi_{\mu} \\
        \sum_{\mu}C^{\mu,\beta}_{p}\chi_{\mu}
\end{pmatrix},
\end{equation}
where $\mathbf{C}$ are expansion coefficients of the molecular spinor, and  $\chi_{\mu}$ are the atomic-centered basis functions.
In this work, we use the density-fitted quasi-second-order two-step optimization algorithm proposed by Hohenstein \textit{et al.}\cite{hohenstein.2015.10.1063/1.4921956} for the alternating optimization of CI coefficients and molecular spinor expansion coefficients.
The optimization of CI coefficients, at fixed spinor coefficients, is performed via a Davidson--Liu iterative eigensolver\cite{davidson.1975.10.1016/0021-99917590065-0,liu.1978.} capable of handling complex Hermitian matrices, together with a complex-arithmetic version of the Harrison--Zarrabian direct CI algorithm,\cite{harrison.1989.10.1016/0009-26148987358-0}, where the product of the Hamiltonian matrix and a trial vector, $\bvec{\sigma}=\bvec{Hc}$, is directly computed, avoiding the explicit and costly construction of the Hamiltonian matrix in the determinant basis.
The spinors are optimized, at fixed CI coefficients, by unitary rotations of the form $\bvec{C}' = \bvec{C} \exp(\bvec{R})$.
In this work, we use a complex-arithmetic L-BFGS algorithm\cite{liu.1989.10.1007/BF01589116} to optimize the spinor rotation parameters contained in the skew-Hermitian matrix $\bvec{R}$ in each macro-iteration, which requires the evaluation of the orbital gradient and diagonal Hessian, whose expressions are adapted from ref~\citenum{chaban.1997.10.1007/s002140050241} for the complex spinor case.
The resulting two-step quasi-second-order algorithm is able to converge X2C-CASSCF wave functions to tight thresholds (10$^{-8}$--10$^{-10}$ $\hartree$) in less than 20 macro-iterations for most systems tested.

The simultaneous orbital optimization for $n$ CI states is possible through the state-averaged (SA) formalism,\cite{werner.1981.10.1063/1.440892} where instead the ensemble average energy is minimized:
\begin{equation}
    \label{eq:sa_energy}
    E_{\text{SA-CASSCF}} = \sum_{k=1}^{n} w_k E_k,
\end{equation}
where $E_k$ is the energy of state $k$, and $w_k$ is the corresponding weight, with $\sum_{k=1}^{n} w_k = 1$.
The only modification to the above algorithm required by the SA formalism is that the density matrices used to compute the orbital gradient and diagonal Hessian are replaced by the weighted average of the density matrices of each state.

Finally, we note that the X2C-CASSCF method as implemented by us does not exploit time-reversal (or Kramers) symmetry, but as has been demonstrated by Kasper \textit{et al.},\cite{kasper.2020.10.1063/5.0015279} the SA-CASSCF orbital optimization procedure can essentially fully recover Kramers degeneracy.

\subsection{X2C-DSRG-MRPT2 and Its Approximations}
\begin{table*}[!htpb]
\caption{Summary of the different schemes for the inclusion of SOC effects. The one-electron Hamiltonian used in each step is indicated for each scheme. ``CI post conv.'' refers to the final CI iteration after the CASSCF optimization has converged, but before the DSRG-MRPT2 energy evaluation step. ``$H_{\mathrm{eff.}}$ diag.'' refers to the diagonalization of the DSRG-MRPT2 effective Hamiltonian in the active space.}
\label{tab:schemes}
\begin{tabular}{lllll}
\hhline{=====}
Scheme                   & CASSCF           & CI post conv.  & DSRG-MRPT2 & $H_{\mathrm{eff.}}$ diag.\\ \hline
sf-X2C-CASSCF-SO         & sf-X2C           & SNSO-X2C        & --        & --              \\
sf-X2C-DSRG-MRPT2-SO (A) & sf-X2C           & sf-X2C          & sf-X2C    & SNSO-X2C        \\
sf-X2C-DSRG-MRPT2-SO (B) & sf-X2C           & sf-X2C          & SNSO-X2C  & SNSO-X2C        \\
sf-X2C-DSRG-MRPT2-SO (C) & sf-X2C           & SNSO-X2C        & SNSO-X2C  & SNSO-X2C        \\
X2C-CASSCF               & SNSO-X2C         & SNSO-X2C        & --        & --              \\
X2C-DSRG-MRPT2           & SNSO-X2C         & SNSO-X2C        & SNSO-X2C  & SNSO-X2C        \\
\hhline{=====}
\end{tabular}
\end{table*}
Since the DSRG-MRPT2 method has already been described in detail in previous works,\cite{li.2015.10.1021/acs.jctc.5b00134} we recapitulate the key equations in Appendix~\labelcref{sec:dsrg} for the interested reader.
The equations presented therein are presented in the spin--orbital basis and can be directly used to generate tensor contractions that work with any complex Hermitian second-quantized Hamiltonians, as shown in our previous work on state-averaged four-component DSRG-MRPT2/3.\cite{zhao.2024.10.1021/acs.jpclett.4c01372}
They can also be efficiently implemented using the density fitting (DF) approximation to avoid the storage of four-index intermediates, whose non-relativistic (real arithmetic) implementations in the MR-DSRG framework have been discussed in detail in previous works.\cite{hannon.2016.10.1063/1.4951684,zhang.2019.10.1021/acs.jctc.9b00353}
In this work, we present the first density-fitted SA-DSRG-MRPT2 implementation in a complex spinor basis, enabling its use with practically any relativistic Hamiltonian.
The X2C-DSRG-MRPT2 method requires at most the three-body reduced density cumulant of the reference wave function, and in the common case of $N_{\mathrm{A}} \ll N_{\mathrm{C}} < N_{\mathrm{V}}$, where $N_{\mathrm{C/A/V}}$ are the number of core, active, and virtual spinors, respectively, it scales asymptotically as $\mathcal{O}(N_{\mathrm{C}}^2 N_{\mathrm{V}}^2)$ (see Appendix~\labelcref{sec:dsrg} for details).
This means that, compared to the non-relativistic counterpart, the prefactor of the X2C-DSRG-MRPT2 method is increased by a factor of 16 due to the use of two-component spinors, and further by a factor of roughly 4 due to the use of complex arithmetic,\cite{dyall.2007.10.1093/oso/9780195140866.001.0001} resulting in an overall prefactor increase of about 64.
The asymptotic scaling with respect to the number of active spinors is $\mathcal{O}(N_{\mathrm{A}}^6N_{\mathrm{V}})$.\cite{li.2015.10.1021/acs.jctc.5b00134}

The X2C-DSRG-MRPT2 is a fully two-component relativistic method that uses the SNSO-X2C Hamiltonian to include relativistic effects, optimizes the molecular spinors \emph{via} the X2C-CASSCF method, and treats dynamic correlation through the DSRG-MRPT2 formalism.
We also present three approximations to the X2C-DSRG-MRPT2 scheme that differ in the point in the calculation at which the X2C Hamiltonian is switched on.
This idea has been explored for other relativistic methods, such as by Liu and coworkers for EOM-CCSD\cite{cao.2017.10.1039/C6CP07588F} and by Li and coworkers for the uncontracted X2C-MRPT2 method.\cite{lu.2022.10.1021/acs.jctc.2c00171}
These schemes are summarized in \cref{tab:schemes}.

In the first scheme, which we term sf-X2C-DSRG-MRPT2-SO (A), the molecular orbitals are optimized with the spin-free one-electron X2C Hamiltonian and a non-relativistic SA-CASSCF computation, followed by a non-relativistic DSRG-MRPT2 calculation, and the SNSO-X2C Hamiltonian is only used in the reference relaxation step after the DSRG-MRPT2 calculation.
In the second scheme, termed sf-X2C-DSRG-MRPT2-SO (B), the SNSO-X2C Hamiltonian is employed in the DSRG-MRPT2 calculation, with the spin-free SA-CASSCF ensemble of states serving as the reference.
The third scheme, termed sf-X2C-DSRG-MRPT2-SO (C), further includes the SNSO-X2C Hamiltonian in a final CI iteration after the CASSCF optimization has converged. The resulting spin--orbit coupled ensemble of states is then used as the reference for the subsequent relativistic DSRG-MRPT2 calculation and reference relaxation.

Also included for comparison are the X2C-CASSCF scheme and what we call the sf-X2C-CASSCF-SO scheme. 
The latter includes one iteration of CI optimization with the SNSO-X2C Hamiltonian after the convergence of a scalar-relativistic SA-CASSCF calculation using the sf-X2C Hamiltonian, but does not include any dynamic correlation treatment, which is essentially the sf-X2C-DSRG-MRPT2-SO (C) scheme without the DSRG-MRPT2 step.

In all four spin-free schemes, orbital optimization is performed without considering vector relativistic effects and proceeds in the spin-free formalism, which significantly reduces the computational cost in both the CI and orbital rotation steps, and in the sf-X2C-DSRG-MRPT2-SO (A) scheme, even the DSRG-MRPT2 energy evaluation step is performed in the spin-free formalism, which further reduces the computational cost.
We note that, while the sf-X2C-DSRG-MRPT2-SO (B) and (C) approximations have not been proposed in existing literature, the sf-X2C-DSRG-MRPT2-SO (A) scheme is essentially identical to the X2CSO-DSRG-MRPT2 method recently proposed by Park,\cite{park.2025.10.1021/acs.jctc.5c01543} and very similar to the BP1-SA-DSRG-PT2 method, which uses the first-order Breit--Pauli (BP1) Hamiltonian with spin--orbit mean-field contributions to account for two-electron SOC effects.\cite{wang.2025.10.3390/molecules30092082}
Nevertheless, as we will show in \cref{sec:pblock}, these approximations can lead to significant loss of accuracy for systems with strong SOC effects.

\section{Computational Details}
\label{sec:comp_details}
The SNSO-X2C Hamiltonian, X2C-CASSCF, X2C-DSRG-MRPT2 and its approximate variants have been implemented in the open-source \textsc{Forte2} package (version 2026.5.1), and all calculations in this work were performed using this software.

The Gaussian finite nucleus model was used for all atoms.\cite{visscher.1997.10.1006/adnd.1997.0751}
The row-dependent SNSO parameters from ref~\citenum{ehrman.2023.10.1021/acs.jctc.3c00479} were used in all calculations, unless otherwise specified.
The speed of light was set to $c=137.035999177$ atomic units, and a conversion factor of $219474.63136314$ \wn/$\hartree$ was used to convert energies from atomic units to wavenumbers.
The density fitting (DF) approximation was used in all X2C-DSRG-MRPT2 calculations, and unless otherwise specified, the auxiliary basis set was generated from the computational basis set using the AutoAux procedure\cite{stoychev.2017.10.1021/acs.jctc.6b01041} as implemented in the \textsc{BasisSetExchange} package.\cite{pritchard.2019.10.1021/acs.jcim.9b00725,lehtola.2021.10.1021/acs.jctc.1c00607}
All computational results are available in the supplementary material.

\section{Results and Discussion}
\label{sec:results}

\subsection{Comparison between Approximate SOC Schemes}

The zero-field splittings (ZFS) of the $p$-block elements are popular benchmarks for the accuracy of relativistic electronic structure methods, as they range from a few \wn~in the second row to tens of thousands \wn~in the sixth row, and hence test both the precision and accuracy of the method.
The accurate calculation of these splittings also requires that a method can balance the treatment of relativistic effects and electron correlation.

To start, we first examine the performance of the approximate schemes introduced in \cref{tab:schemes} for the treatment of SOC, comparing them with the X2C-DSRG-MRPT2 method, which employs a fully variational treatment of SOC at the X2C-CASSCF level.
Here we compute the zero-field splittings of second- to sixth-row $p$-block elements, and compare them to experimental values.
The decontracted ANO-RCC basis set was used for all atoms,\cite{widmark.1990.10.1007/BF01120130,roos.2004.10.1021/jp031064+} and the $n$s $n$p active space was used for all second- to fifth-row elements, while the $n$p $(n+1)$s active space was used for the sixth-row elements, where $n$ is the row number of the element.
State averaging was performed for the lowest 6, 9, 14, 9, and 6 roots for groups 13 to 17, respectively, with equal weights assigned to all states.
Here and elsewhere, unless otherwise stated, a flow parameter of $s=0.50$ \sunit~was used for all MR-DSRG computations, and all electrons were correlated in all orbitals.

\begin{figure}[!htbp]
    \centering
    \includegraphics[width=3.25in]{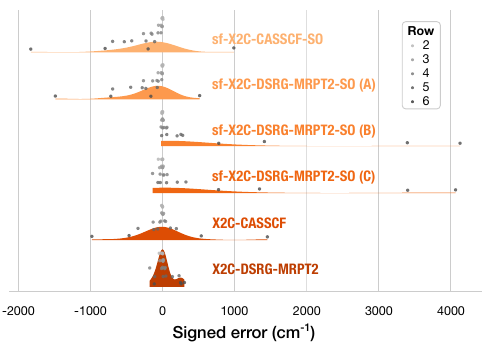}
    \caption{Spin--orbit splittings of the second- to sixth-row $p$-block elements.    
    Comparison of the signed error (in \wn) with respect to experimental values of the five different treatments of spin--orbit coupling based on X2C-DSRG-MRPT2. All results were obtained with a decontracted ANO-RCC basis and $s$ = 0.5~\sunit.}
    \label{fig:so_compare}
\end{figure}

In \cref{fig:so_compare}, we show the errors of all schemes in \cref{tab:schemes} compared to experimental values, broken down by row of the periodic table.
These data show relatively good performance of all the approximate schemes persisting into the fourth row, but with a substantial increase in error for the fifth- and sixth-row elements, especially for the sf-X2C-DSRG-MRPT2-SO (B) and (C) schemes for the sixth row, but all approximate schemes predict essentially qualitatively wrong splittings for the sixth-row elements.
The X2C-DSRG-MRPT2 method, with its fully variational treatment of SOC at the X2C-CASSCF level, maintains the same level of accuracy in terms of MAPE for the heavier elements as for the lighter elements, while the other four schemes show a substantial increase in MAPE for the heavier elements.
Comparing the X2C-CASSCF and X2C-DSRG-MRPT2 results also shows directly that incorporating dynamical correlation is essential for accurate spin--orbit splittings.
We provide a detailed analysis to this end in \cref{sec:dynamical_correlation}.

A subtler point, however, concerns the relative performance of the four approximate schemes.
The SNSO-X2C Hamiltonian enters into the computation in the order (A) $\rightarrow$ (B) $\rightarrow$ (C), and one might expect that the performance of the schemes would improve in this order, but this is evidently not the case.
In fact, the sf-X2C-CASSCF-SO and sf-X2C-DSRG-MRPT2-SO (A) schemes share very similar, and better, error profiles, while the (B) and (C) schemes share another similar, but worse, error profile.
This suggests that the main difference between these two sets is whether DSRG-MRPT2 amplitudes are computed in the presence of the SNSO-X2C Hamiltonian, with those that do not (sf-X2C-CASSCF-SO and sf-X2C-DSRG-MRPT2-SO (A)) performing better than those that do (sf-X2C-DSRG-MRPT2-SO (B) and (C)).

A further interesting point can be observed: the spread of errors in X2C-CASSCF is similar to sf-X2C-CASSCF-SO, but the spread of errors in X2C-DSRG-MRPT2 is much smaller than in sf-X2C-DSRG-MRPT2-SO (A).
This suggests that the inclusion of orbital relaxation \emph{alone} is not sufficient to achieve the best accuracy, and that the inclusion of \emph{consistent} dynamical correlation on top of orbital relaxation is necessary.

\subsection{Comparison to Four-Component Methods}
Having established the importance of treating SOC variationally at the X2C-CASSCF level for accurate spin--orbit splittings, we now focus only on the X2C-DSRG-MRPT2 method.
In this section, we benchmark the performance of X2C-DSRG-MRPT2 for the zero-field splittings of the second- to fourth-row $p$-block elements, comparing to our previous four-component DSRG-MRPT2/3 results,\cite{zhao.2024.10.1021/acs.jpclett.4c01372} as well as other four-component methods from the literature.
In this section only, we use a flow parameter of $s=0.24$ \sunit~here for the DSRG-MRPT2 computations, to compare to 4C-DSRG-MRPT2 in our previous work, as this value was shown to give the best agreement with experiment for these systems in that work.
To enable one-to-one comparison with published data, the decontracted cc-pVTZ basis set was used for all atoms,\cite{dunning.1989.10.1063/1.456153,wilson.1999.10.1063/1.478678,woon.1993.10.1063/1.464303} while the active space was chosen to be the full valence space ($n$s, $n$p) and all non-valence spinors were frozen for third- and fourth-row atoms in all correlated computations.
While unnecessary for X2C-DSRG-MRPT2, this freezing scheme was introduced for the four-component methods, both for convergence and computational tractability.
State-averaging was performed for the lowest 6, 9, 14, 9, and 6 states for groups 13 to 17, respectively, with equal weights assigned to all states.
Unless otherwise specified, all experimental atomic splittings in this subsection are taken from the NIST Atomic Spectra Database.\cite{kramida.2024.10.18434/T4W30F}

\begin{table*}[!htbp]
\scriptsize
\begin{threeparttable}
\caption{Comparison between the spin--orbit splittings of the 15 second- to fourth-row $p$-block elements computed with X2C-CASSCF, X2C-DSRG-MRPT2, 4C-SA-CASSCF, 4C-CASPT2, 4C-MR-CISD+Q, 4C-MRPT2 and 3, and 4C-iCIPT2 to the experimental splittings. All results are reported in units of \wn{} and use the decontracted cc-pVTZ basis set, unless otherwise noted. ``Exp.'' stands for experimental splitting, ``MAE'' stands for mean absolute error, and ``MAPE'' stands for mean absolute percentage error. All four-component methods used the Dirac--Coulomb--Breit Hamiltonian. The 4C-DSRG-MRPT2/3 results are taken from ref~\citenum{zhao.2024.10.1021/acs.jpclett.4c01372}, and the 4C-iCIPT2 results are from ref~\citenum{zhang.2024.10.1021/acs.jctc.4c00967}, while the remaining four-component results are taken from ref~\citenum{zhang.2018.10.1021/acs.jctc.7b00989}. Results from the current work (X2C-CASSCF and X2C-DSRG-MRPT2) are presented in boldface here and elsewhere.}
\label{tab:pblock_tz}
\begin{tabular}{lBBlllllll}
\hhline{==========}
Splitting 
& X2C- & X2C-DSRG- & 4C- & 4C-DSRG- & 4C-DSRG- & 4C- & 4C-MR- & 4C- & Exp. \\
& CASSCF 
& MRPT2\tnote{e}
& CASSCF
& MRPT2\tnote{e}
& MRPT3\tnote{e}
& CASPT2
& CISD+Q
& iCIPT2
& \\ \hline

B $^2$P$_{\sfrac{1}{2}}\rightarrow^2$P$_{\sfrac{3}{2}}$
& 13.45 & 14.01 & 13.25 & 14.25 & 14.32 & 13.99 & 13.91 & 13.97 & 15.29 \\

C $^3$P$_{0}\rightarrow^3$P$_{1}$
& 14.80 & 16.41 & 14.94 & 16.96 & 17.20 & 14.93 & 15.40 & 15.61 & 16.42 \\

N $^2$D$_{\sfrac{5}{2}}\rightarrow^2$D$_{\sfrac{3}{2}}$
& -2.05 & -4.42 & 11.09 & 8.16 & 7.89 & 9.41 & 9.40\tnote{a} & 9.54\tnote{a} & 8.71 \\

O $^3$P$_{2}\rightarrow^3$P$_{1}$
& 142.11 & 115.71 & 153.23 & 129.99 & 138.92 & 145.35 & 152.52 & 152.08 & 158.26 \\

F $^2$P$_{\sfrac{3}{2}}\rightarrow^2$P$_{\sfrac{1}{2}}$
& 393.71 & 389.45 & 382.58 & 380.46 & 391.68 & 384.70 & 388.38 & 387.66 & 404.14 \\ \hline

Al $^2$P$_{\sfrac{1}{2}}\rightarrow^2$P$_{\sfrac{3}{2}}$
& 98.40 & 106.16 & 96.81 & 106.90 & 108.35 & 106.70 & 106.96 & 106.43 & 112.06 \\

Si $^3$P$_{0}\rightarrow^3$P$_{1}$
& 73.72 & 78.81 & 72.89 & 78.90 & 81.52 & 69.94 & 73.76 & 73.80 & 77.11 \\

P $^2$D$_{\sfrac{3}{2}}\rightarrow^2$D$_{\sfrac{5}{2}}$
& 18.25 & 16.62 & 15.34 & 14.01 & 13.38 & 11.08 & --\tnote{b} & 13.62 & 15.61 \\

S $^3$P$_{2}\rightarrow^3$P$_{1}$
& 399.48 & 383.94 & 398.64 & 386.02 & 400.61 & 355.94 & 383.94 & 386.11 & 396.06 \\

Cl $^2$P$_{\sfrac{3}{2}}\rightarrow^2$P$_{\sfrac{1}{2}}$
& 894.20 & 896.56 & 886.86 & 894.86 & 888.44 & 867.69 & 861.80 & 864.17 & 882.35 \\ \hline

Ga $^2$P$_{\sfrac{1}{2}}\rightarrow^2$P$_{\sfrac{3}{2}}$
& 695.48 & 746.20 & 685.92 & 776.51 & 791.62 & 743.28 & 745.97 & 739.16 & 826.19 \\

Ge $^3$P$_{0}\rightarrow^3$P$_{1}$
& 518.65 & 547.98 & 512.35 & 553.07 & 570.25 & 485.56 & 502.94\tnote{c} & 504.81 & 557.13 \\

As $^2$D$_{\sfrac{3}{2}}\rightarrow^2$D$_{\sfrac{5}{2}}$
& 362.17 & 331.75 & 354.53 & 324.93 & 327.81 & 227.88 & --\tnote{b} & 285.05 & 322.10 \\

Se $^3$P$_{2}\rightarrow^3$P$_{1}$
& 1963.98 & 1913.43 & 1949.63 & 1917.35 & 1991.36 & 1745.74 & 1900.34\tnote{c} & 1886.09 & 1989.50 \\

Br $^2$P$_{\sfrac{3}{2}}\rightarrow^2$P$_{\sfrac{1}{2}}$
& 3708.96 & 3693.50 & 3683.62 & 3704.50 & 3683.90 & 3546.46 & 3540.14 & 3548.41 & 3685.24 \\ \hline

MAE 
& 22.3 & 19.3 & 21.2 & 15.5 & 7.5 & 49.3 & 33.4\tnote{d} & 32.1 & -- \\

MAPE\tnote{f}
& 7.7\% & 5.4\% & 7.8\% & 4.9\% & 4.6\% & 10.8\% & 5.6\%\tnote{d} & 6.5\% & -- \\
\hhline{==========}
\end{tabular}
\begin{tablenotes}
\scriptsize
\item [a] Two core spinors were frozen for nitrogen.
\item [b] The 4C-MRCISD+Q computations were intractable for these atoms.
\item [c] The decontracted cc-pVDZ basis set was used for these atoms.
\item [d] Unavailable data points have been omitted from averaging.
\item [e] A flow parameter of $s=0.24$ \sunit~was used for all DSRG-MRPT2 computations, and $s=0.35$ \sunit~was used for all DSRG-MRPT3 computations.
\item [f] The percentage error for the nitrogen splitting is omitted from the MAPE calculation, as its percentage error is anomalously large for some methods due to the small magnitude of the splitting.
\end{tablenotes}
\end{threeparttable}
\end{table*}

As shown in \cref{tab:pblock_tz}, X2C-DSRG-MRPT2 yields results that are in excellent agreement with the much more expensive four-component equivalent, with a mean absolute error (MAE) of 19.3 \wn~and a mean absolute percentage error (MAPE) of 5.4\%, compared to 15.5 and 4.9\% for 4C-DSRG-MRPT2.
It is also noteworthy that the X2C-DSRG-MRPT2 is likewise able to improve upon the X2C-CASSCF results, reducing the MAE from 22.3 to 19.3 \wn~and the MAPE from 7.7\% to 5.4\%, again following the trend seen in their four-component analogs.
This shows that the SNSO-X2C Hamiltonian is able to accurately reproduce the many-electron spectrum of the four-component Dirac--Coulomb--Breit Hamiltonian, which is consistent with previous studies.\cite{zhang.2024.10.1021/acs.jpca.3c08167,yuwono.2025.10.1063/5.0248535}
We also note that the 4C-iCIPT2 results can be considered to be near the FCI limit for a given one-particle basis set,\cite{zhang.2024.10.1021/acs.jctc.4c00967} and the fact that X2C-DSRG-MRPT2 is able to achieve a similar level of accuracy to 4C-iCIPT2 shows that the DSRG-MRPT2 method is able to capture the essential dynamical correlation effects for these systems.

Deserving further attention is the comparison between the X2C-DSRG-MRPT2 and 4C-DSRG-MRPT2 results.
For the present set of 15 splittings, the two methods share identical computational parameters, and hence the only difference between the two methods is the use of the SNSO-X2C Hamiltonian in the X2C-DSRG-MRPT2 method, and the use of the four-component Dirac--Coulomb--Breit Hamiltonian in the 4C-DSRG-MRPT2 method.
Despite the large computational savings enabled by the use of the SNSO-X2C Hamiltonian (together with an efficient density-fitted implementation), the two methods show excellent agreement, with a regression analysis of the error with respect to experiment showing a correlation coefficient of $0.9497$, and the MAE between the two methods is only $6.73$ \wn, meaning their results are well correlated.
This justifies the use of the SNSO-X2C Hamiltonian as a reliable and efficient approximation to the parent four-component Dirac--Coulomb--Breit Hamiltonian for the treatment of relativistic effects.
Notable outliers, such as nitrogen and oxygen, exhibit large percentage errors due to the small magnitude of their splittings.
These outliers, along with the regression analysis, reveal that the SNSO-X2C Hamiltonian has an intrinsic error of a few wavenumbers for these splittings, and as such might not be able to provide a systematic path to spectroscopic accuracy (errors of 1 \wn~or better) for these systems, even when paired with an accurate treatment of electron correlation.

\subsection{Zero-Field Splitting of \textit{p}-Block Elements}
\label{sec:pblock}
\begin{table*}[htpb]
\scriptsize
\begin{threeparttable}
\caption{The experimental and computed zero-field splittings (in \wn) of the second- to fifth-row $p$-block elements, as well as thallium. The EOM-CCSD(SOC) results are taken from ref~\citenum{cao.2017.10.1039/C6CP07588F}; the SO-EOM-CCSD results are from ref~\citenum{cheng.2018.10.1063/1.5012041}; the BP1-DSRG-MRPT2 results are from ref~\citenum{wang.2025.10.3390/molecules30092082}; the BP1- and DKH2-NEVPT2 results are from ref~\citenum{majumder.2024.10.1021/acs.jctc.4c00458}; the X2C-MC-PDFT results are from ref~\citenum{sharma.2022.10.1021/acs.jctc.2c00062}; and the 4C-iCIPT2 results are from ref~\citenum{zhang.2024.10.1021/acs.jctc.4c00967}.}
\label{tab:pblock_ano}
\begin{tabular}{llllllBlBll}
\hhline{===========}
Splitting 
& EOM-CCSD & SO-EOM- & BP1-DSRG- & BP1- & DKH2- & X2C- & X2C- & X2C-DSRG- & 4C- & Exp. \\
& (SOC)\tnote{c} 
& CCSD 
& MRPT2\tnote{b} 
& NEVPT2 
& NEVPT2 
& CASSCF 
& MC-PDFT 
& MRPT2\tnote{b} 
& iCIPT2\tnote{a}
& \\ \hline

B $^2$P$_{\sfrac{1}{2}}\rightarrow^2$P$_{\sfrac{3}{2}}$
& 13.7 & 13.7 & 14.9 & 15.0 & 14.5 & 13.70 & -- & 14.23 & 19.20 & 15.29 \\

C $^3$P$_{0}\rightarrow^3$P$_{1}$
& -- & -- & -- & -- & -- & 15.08 & -- & 17.71 & 18.54 & 16.42 \\

N $^2$D$_{\sfrac{5}{2}}\rightarrow^2$D$_{\sfrac{3}{2}}$
& -- & -- & -- & -- & -- & -2.15 & -- & -6.58 & -3.43\tnote{f} & 8.71 \\

O $^3$P$_{2}\rightarrow^3$P$_{1}$
& -- & -- & -- & -- & -- & 144.63 & -- & 110.28 & 160.30 & 158.26 \\

F $^2$P$_{\sfrac{3}{2}}\rightarrow^2$P$_{\sfrac{1}{2}}$
& 397.7 & 397.7 & 401.5 & 401.5 & 405.0 & 401.32 & -- & 395.53 & 438.00 & 404.14 \\ \hline

Al $^2$P$_{\sfrac{1}{2}}\rightarrow^2$P$_{\sfrac{3}{2}}$
& 107.7 & 107.5 & 105.8 & 107.6 & 109.4 & 98.28 & -- & 114.80 & 113.90 & 112.06 \\ 

Si $^3$P$_{0}\rightarrow^3$P$_{1}$
& -- & -- & -- & -- & -- & 73.75 & -- & 82.53 & 78.32 & 77.11 \\

P $^2$D$_{\sfrac{3}{2}}\rightarrow^2$D$_{\sfrac{5}{2}}$
& -- & -- & -- & -- & -- & 18.53 & -- & 17.98 & 18.57 & 15.61 \\

S $^3$P$_{2}\rightarrow^3$P$_{1}$
& -- & -- & -- & -- & -- & 399.46 & -- & 356.99 & 402.40 & 396.06 \\

Cl $^2$P$_{\sfrac{3}{2}}\rightarrow^2$P$_{\sfrac{1}{2}}$
& 876.0 & 872.8 & 789.7 & 789.7 & 858.6 & 895.27 & -- & 894.71 & 902.10 & 882.35 \\ \hline

Ga $^2$P$_{\sfrac{1}{2}}\rightarrow^2$P$_{\sfrac{3}{2}}$
& 812.9 & 797.6 & 865.7 & 887.4 & 818.8 & 694.19 & 780.0 & 861.24 & 755.70 & 826.19 \\

Ge $^3$P$_{0}\rightarrow^3$P$_{1}$
& -- & -- & -- & -- & -- & 518.31 & -- & 586.32 & 518.80 & 557.13 \\

As $^2$D$_{\sfrac{3}{2}}\rightarrow^2$D$_{\sfrac{5}{2}}$
& -- & -- & -- & -- & -- & 363.23 & 310.0 & 307.55 & 301.30 & 322.10 \\

Se $^3$P$_{2}\rightarrow^3$P$_{1}$
& -- & -- & -- & -- & -- & 1964.04 & -- & 1810.16 & 1929.00 & 1989.50 \\

Br $^2$P$_{\sfrac{3}{2}}\rightarrow^2$P$_{\sfrac{1}{2}}$
& 3648.8 & 3555.4 & 3574.4 & 3574.4 & 3625.0 & 3712.04 & 3930.0 & 3684.28 & 3616.00 & 3685.24 \\ \hline

In $^2$P$_{\sfrac{1}{2}}\rightarrow^2$P$_{\sfrac{3}{2}}$
& 2214.8 & 2103.6 & 2470.5 & 2560.8 & 2219.0 & 1870.42 & 2160.0 & 2347.97 & 1989.00 & 2212.60 \\

Sn $^3$P$_{0}\rightarrow^3$P$_{1}$
& -- & -- & -- & -- & -- & 1569.35 & 1760.0 & 1736.31 & 1528.00 & 1691.81 \\

Sb $^2$D$_{\sfrac{3}{2}}\rightarrow^2$D$_{\sfrac{5}{2}}$
& -- & -- & -- & -- & -- & 1531.84 & 1350.0 & 1348.93 & 1227.00 & 1341.89 \\

Te $^3$P$_{2}\rightarrow^3$P$_{1}$
& -- & -- & -- & -- & -- & 4638.65 & 4480.0 & 4596.89 & 4536.00 & 4706.49 \\

I $^2$P$_{\sfrac{3}{2}}\rightarrow^2$P$_{\sfrac{1}{2}}$
& 7754.60 & 7288.80 & 8150.1 & 8149.9 & 7581.0 & 7710.82 & 7920.0 & 7830.58 & 7370.00 & 7602.98 \\ \hline

Tl $^2$P$_{\sfrac{1}{2}}\rightarrow^2$P$_{\sfrac{3}{2}}$
& 8210.30 & 6794.10 & 12065.6 & 12475.8 & 8113.30 & 6810.87 & 8299.0 & 8049.64 & -- & 7792.70 \\ \hline

MAE  
& 71.1\tnote{g} & 178.0\tnote{g} & 592.2\tnote{g} & 650.0\tnote{g} & 49.4\tnote{g} & 102.0 & 164.6 & 56.0 & 62.6\tnote{g} & -- \\ 

MAPE 
& 3.0\%\tnote{g} & 5.1\%\tnote{g} & 11.2\%\tnote{g} & 12.3\%\tnote{g} & 2.0\%\tnote{g} & 7.8\%\tnote{h} & 4.3\% & 6.2\%\tnote{h} & 13.8\%\tnote{g,h} & -- \\
\hhline{===========}
\end{tabular}
\begin{tablenotes}
\scriptsize
\item [a] The Dirac--Coulomb Hamiltonian was used for the 4C-iCIPT2 computations. Only valence electrons were correlated.
\item [b] A flow parameter of $s=0.5$ \sunit~was used for the all DSRG computations.
\item [c] The decontracted ANO-RCC-VTZP basis set was used for the EOM-CCSD(SOC) computations.
\item [d] The contracted X2C-TZPall-2c basis set was used for the X2C-MRCISD computations.
\item [e] The decontracted X2C-TZVPall basis set was used for the X2CSO-DSRG-MRPT2 computations, with the decontracted X2C-QZVPall basis set being used for the auxiliary basis set.
\item [f] A negative splitting indicates that the ordering of the $^2$D$_{3/2}$ and $^2$D$_{5/2}$ states is flipped.
\item [g] Unavailable data points have been omitted from averaging.
\item [h] The percentage error for the nitrogen splitting is omitted from the MAPE computation, as its percentage error is anomalously large due to the small magnitude of the splitting.
\end{tablenotes}
\end{threeparttable}
\end{table*}

\begin{figure}[!htbp]
    \centering
    \includegraphics[width=3.25in]{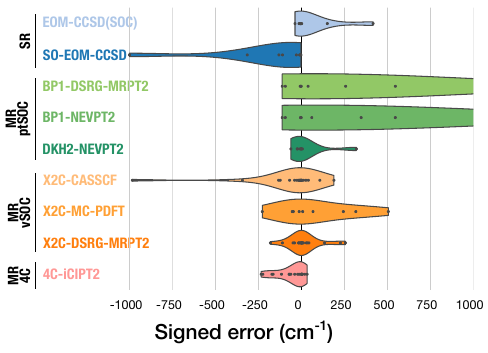}
    \caption{Spin--orbit splittings of the second- to fifth-row $p$-block elements.    
    Comparison of the signed errors (in \wn) with respect to experimental values of various methods. All results were obtained with a decontracted ANO-RCC basis and $s$ = 0.5~\sunit.}
    \label{fig:pblock_ano}
\end{figure}

The decontracted cc-pVTZ basis set used above is relatively modest, and in any case cannot be considered to approach the complete basis set (CBS) limit, and as such, the comparisons to experimental values may not be warranted.
To address this limitation, we next perform computations with the decontracted ANO-RCC basis set,\cite{widmark.1990.10.1007/BF01120130,roos.2004.10.1021/jp031064+} which is close to the CBS limit.
For these computations, we additionally consider the fifth and sixth row $p$-block elements, and compare to other theoretical results from the literature, as shown in \cref{tab:pblock_ano,fig:pblock_ano}.
Some results were computed with different basis sets, as noted in the table footnotes, but all can be considered to approach the atomic CBS limit, and hence can be reasonably compared to experiment. 

These results further corroborate the accuracy of the X2C-DSRG-MRPT2 method: it achieves an MAE of 56.0 \wn, the lowest among all methods compared for the full 21-splitting set, including even the 4C-iCIPT2 method.
The MAPE of 6.2\% is also among the lowest of all methods, except for DKH2-NEVPT2, EOM-CCSD(SOC), and SO-EOM-CCSD, for which fewer data points are available.
The same trend of X2C-DSRG-MRPT2 improving upon X2C-CASSCF is also observed, with the MAE decreasing from 102.0 to 56.0 \wn, and the MAPE decreasing from 7.8\% to 6.2\%.
It is worth noting that the SNSO-X2C Hamiltonian appears to maintain its accuracy even for the heavier elements, unlike the BP Hamiltonian, which is considered a low-$Z$ approximation, due to its reliance on a poorly convergent perturbative expansion of the Dirac Hamiltonian.
On the other hand, the DKH2 Hamiltonian performs well for these systems, as it treats relativistic effects with a convergent series of unitary transformations, whose infinite-order limit is the X2C Hamiltonian.\cite{kedziera.2004.10.1063/1.1792131,li.2012.10.1063/1.4758987,li.2014.10.1063/1.4891567}

\begin{table*}[!htbp]
\tiny
\begin{threeparttable}
\caption{Experimental and computed zero-field splittings (in \wn) of the sixth-row $p$-block elements. The values in parentheses are the absolute percentage errors with respect to experiment, where available. The DKH1-SO-L-PDFT results are taken from ref~\citenum{jangid.2025.10.1021/acs.jctc.5c01633}; the X2C-MC-PDFT results are from ref~\citenum{sharma.2022.10.1021/acs.jctc.2c00062}; the EOM-CCSD(SOC) results are from ref~\citenum{cao.2017.10.1039/C6CP07588F}; the BP1- and DKH2-NEVPT2 results are from ref~\citenum{majumder.2024.10.1021/acs.jctc.4c00458}; the X2CSO-DSRG-MRPT2 results are from ref~\citenum{park.2025.10.1021/acs.jctc.5c01543}.}
\label{tab:sixth_row}
\begin{tabular}{lllllllBBl}
\hhline{==========}
Splitting 
& DKH1-SO- & X2C- & EOM-CCSD & BP1- & DKH2- & X2CSO-DSRG- & X2C- & X2C-DSRG- & Exp. \\
& L-PDFT\tnote{a}
& MC-PDFT
& (SOC)\tnote{b}
& NEVPT2\tnote{d}
& NEVPT2\tnote{d}
& MRPT2\tnote{c}
& CASSCF
& MRPT2
& \\ \hline

Tl $^2$P$_{\sfrac{1}{2}}\rightarrow^2$P$_{\sfrac{3}{2}}$
& 7717 (0.97)
& 8299 (6.50)
& 8210.3 (5.36)
& 12475.8 (60.1)
& 8113.3 (4.11)
& 8332 (6.92)
& 6810.87 (12.60)
& 8049.64 (3.30)
& 7792.7 \\

Pb $^3$P$_{0}\rightarrow^3$P$_{1}$
& 6300 (19.4)
& 6970 (10.9)
& --
& --
& --
& --
& 7353.14 (5.96)
& 7702.56 (1.49)
& 7819.2626 \\

Bi $^2$D$_{\sfrac{3}{2}}\rightarrow^2$D$_{\sfrac{5}{2}}$
& 3900 (2.95)
& 3940 (1.95)
& --
& --
& --
& --
& 4554.22 (13.33)
& 4265.80 (6.16)
& 4018.462 \\

Po $^3$P$_{2}\rightarrow^3$P$_{0}$
& 6800 (9.51)
& 5940 (21.0)
& --
& --
& --
& --
& 8967.56 (19.33)
& 7816.65 (4.02)
& 7514.69 \\

At $^2$P$_{\sfrac{3}{2}}\rightarrow^2$P$_{\sfrac{1}{2}}$
& 19962
& 19580
& 24880.5
& 34153.5
& 23002.4
& 20480
& 23032.04
& 21338.29
& -- \\ \hline

MAE
& 670.0
& 752.2
& --\tnote{e}
& --\tnote{e}
& --\tnote{e}
& --\tnote{e}
& 859.1
& 230.7
& -- \\

MAPE
& 8.21\%
& 10.1\%
& --\tnote{e}
& --\tnote{e}
& --\tnote{e}
& --\tnote{e}
& 12.8\%
& 3.74\%
& -- \\
\hhline{==========}
\end{tabular}
\begin{tablenotes}
\scriptsize
\item [a] The ANO-RCC-VTZP basis set, and tPBE0 on-top functional were used for the DKH1-SO-L-PDFT computations.
\item [b] The decontracted ANO-RCC-VTZP basis set was used for the EOM-CCSD(SOC) computations.
\item [c] The decontracted X2C-TZVPall basis set was used for the X2CSO-DSRG-MRPT2 computations, with the decontracted X2C-QZVPall basis set being used for the auxiliary basis set. A flow parameter of $s=0.5$ \sunit~was used for these computations.
\item [d] The decontracted ANO-RCC-VTZP basis set was used for the NEVPT2 computations.
\item [e] MAE and MAPE are omitted due to insufficient data points.
\end{tablenotes}
\end{threeparttable}
\end{table*}

Finally, we push our computations to the heaviest $p$-block elements for which experimental data is available.
In \cref{tab:sixth_row}, we compare our X2C-DSRG-MRPT2 results to experimental data (unavailable for astatine) as well as other theoretical results from the literature.
The X2C-DSRG-MRPT2 method continues to perform well.
Although the MAE is now more substantial at 230.7 \wn, these correspond to an MAPE of 3.74\%, which is even somewhat lower than that for the lighter elements.
The X2C-DSRG-MRPT2 method also outperforms the DKH1-SO-L-PDFT method in both MAE and MAPE.
As for the ZFS of astatine, X2C-DSRG-MRPT2 agrees very well with DKH2-NEVPT2, which has performed well for the other heavy $p$-block elements.\cite{majumder.2024.10.1021/acs.jctc.4c00458}
The X2C-DSRG-MRPT2 result falls in the middle of the range of values (19580--24880.5 \wn) predicted by the other methods under comparison, excluding BP1-NEVPT2, which significantly overestimates the ZFS of systems containing heavy elements.

\subsection{Zero-Field Splitting of Transition Metal Atoms}
\label{sec:tm}
We next apply the X2C-DSRG-MRPT2 method to the spin--orbit splittings of a set of transition metal atoms.
These atoms are expected to be more strongly correlated than the $p$-block atoms considered above, as the spatially more compact and partially-filled $d$-shells of these atoms can give rise to a dense manifold of quasi-degenerate electronic states, which can be strongly mixed by SOC.
In \cref{tab:tm}, we compute the $^2$D$_{5/2}\rightarrow^2$D$_{3/2}$ splittings of the Cu, Ag, and Au atoms in their first excited states ($nd^9(n+1)s^2$), as well as the $^2$D$_{3/2}\rightarrow^2$D$_{5/2}$ splittings of Sc, Y, and La atoms in their ground states ($nd^1(n+1)s^2$).
Active spaces of 11 electrons in 12 spinors and 3 electrons in 12 spinors were used for the $nd^9(n+1)s^2$ and $nd^1(n+1)s^2$ configurations, respectively, and state-averaging over the lowest 12 and 10 roots was performed for the two configurations, corresponding to the $^2$S$_{1/2}$, $^2$D$_{3/2}$, and $^2$D$_{5/2}$ states of the $nd^9(n+1)s^2$ configurations, and the $^2$D$_{3/2}$ and $^2$D$_{5/2}$ states of the $nd^1(n+1)s^2$ configurations.
The contracted X2C-TZPall-2c basis set\cite{pollak.2017.10.1021/acs.jctc.7b00593} was used for all computations unless otherwise noted.
\begin{table*}[!htbp]
\scriptsize
\caption{The experimental and computed zero-field splittings (in \wn) of selected transition metal atoms. The BP1-DSRG-MRPT2 results are taken from ref~\citenum{wang.2025.10.3390/molecules30092082}, the BP1- and DKH2-NEVPT2 results are from ref~\citenum{majumder.2024.10.1021/acs.jctc.4c00458}, the X2CSO-DSRG-MRPT2 results are from ref~\citenum{park.2025.10.1021/acs.jctc.5c01543}, the DKH1-SO-L-PDFT results are from ref~\citenum{jangid.2025.10.1021/acs.jctc.5c01633}; the X2C-MRCISD results are from ref~\citenum{hu.2020.10.1021/acs.jctc.9b01290}.}
\label{tab:tm}
\begin{threeparttable}
\begin{tabular}{llllllBlBl}
\hhline{==========}
Splitting
& BP1-DSRG-
& BP1-
& DKH2-
& X2CSO-DSRG-
& DKH1-SO-
& X2C-
& X2C-
& X2C-DSRG-
& Exp. \\
& MRPT2\tnote{1}
& NEVPT2
& NEVPT2
& MRPT2\tnote{1,2}
& L-PDFT
& CASSCF
& MRCISD
& MRPT2\tnote{1}
& \\\hline

Cu $^2$D$_{\sfrac{5}{2}}\rightarrow^2$D$_{\sfrac{3}{2}}$
& 2100.00
& --
& --
& 2170.00
& 2090.00
& 2102.16
& --
& 1929.05
& 2042.83 \\

Ag $^2$D$_{\sfrac{5}{2}}\rightarrow^2$D$_{\sfrac{3}{2}}$
& 4700.00
& 4370.00
& 4360.00
& 4620.00
& 4510.00
& 4469.51
& --
& 4481.12
& 4471.93 \\

Au $^2$D$_{\sfrac{5}{2}}\rightarrow^2$D$_{\sfrac{3}{2}}$
& 13900.00
& 13200.00
& 12250.00
& 12920.00
& 12900.00
& 12112.25
& --
& 12613.82
& 12274.01 \\

Sc $^2$D$_{\sfrac{3}{2}}\rightarrow^2$D$_{\sfrac{5}{2}}$
& --
& 174.30
& 140.90
& 220.00
& 167.00
& 147.96
& 190.00
& 152.71
& 168.34 \\

Y $^2$D$_{\sfrac{3}{2}}\rightarrow^2$D$_{\sfrac{5}{2}}$
& --
& 494.20
& 428.40
& 549.00
& 492.00
& 404.40
& 520.00
& 503.86
& 530.35 \\

La $^2$D$_{\sfrac{3}{2}}\rightarrow^2$D$_{\sfrac{5}{2}}$
& --
& 999.90
& 896.60
& 944.00
& 838.00
& 694.10
& 936.00
& 981.20
& 1053.16 \\\hline

MAE
& 637.1\tnote{3}
& 224.7\tnote{3}
& 84.4\tnote{3}
& 183.5
& 161.0
& 121.5
& 49.7\tnote{3}
& 96.1
& -- \\

MAPE
& 7.0\%\tnote{3}
& 5.0\%\tnote{3}
& 10.6\%\tnote{3}
& 9.9\%
& 6.1\%
& 12.4\%
& 8.6\%\tnote{3}
& 4.9\%
& -- \\
\hhline{==========}
\end{tabular}
\begin{tablenotes}
\scriptsize
\item [1] A flow parameter of $s=0.5$ \sunit~was used for all DSRG computations.
\item [2] The decontracted X2C-TZVPall basis set was used for the X2CSO-DSRG-MRPT2 computations, with the decontracted X2C-QZVPall basis set being used for the auxiliary basis set.
\item [3] Unavailable data points have been omitted from averaging.
\end{tablenotes}
\end{threeparttable}
\end{table*}

Here again we can see that the X2C-DSRG-MRPT2 method is among the most accurate methods for these systems, with a MAE of 96.1 \wn~and a MAPE of 4.9\%, which are comparable to the accuracy of DKH2-NEVPT2, as well as the uncontracted X2C-MRCISD method, which performs a MR-RASCI computation in a larger active space (albeit partitioned into several restricted active spaces), and is more expensive than X2C-DSRG-MRPT2.\cite{hu.2020.10.1021/acs.jctc.9b01290}
We also emphasize that the X2C-DSRG-MRPT2 method always uses the smallest active space required to capture the static correlation of the system, and performs state-averaging only over the states of interest, whereas methods based on the state-interaction approach typically require larger active spaces and state-averaging over a larger number of states to achieve similar accuracy.
For example, for the $nd^1(n+1)s^2$ configurations, the NEVPT2 computations of ref~\citenum{majumder.2024.10.1021/acs.jctc.4c00458} used an active space of 3 electrons in 9 spatial orbitals (equivalent to 18 spinors), corresponding to the $nd$, $(n+1)s$, and $(n+1)p$ orbitals, and a large number of states were included in the state-averaging -- the equivalent of 38, 98, and 80 states were included for Sc, Y, and La, respectively -- to achieve the reported accuracy.
Admittedly, since these computations were performed with the spin-free CASSCF method, the cost is certainly manageable and lower than if the equivalent state-averaged X2C-CASSCF computations were performed. 
However, since the active space and number of states included become almost empirical parameters, and do not necessarily follow straightforwardly from chemical intuition, the state-interaction approach can be more difficult to apply in a black-box manner.

\subsection{Zero-Field Splitting of Diatomics}
\label{sec:diatomics}
Having established the accuracy of the X2C-DSRG-MRPT2 method for atomic ZFS computations, we next apply the method to the spin--orbit splittings of a set of open-shell diatomic molecules containing up to fifth-row elements.
These molecules in their doublet ground states near their equilibrium geometries are not expected to be very strongly correlated, especially for those containing lighter elements.
However, the presence of strong SOC can lead to significant mixing of the low-lying electronic states, and when excited states with double excitation character are requested, single-reference methods may struggle to provide a balanced description of these states.
\begin{table*}[!htbp]
\scriptsize
\caption{The computed and experimental spin--orbit splittings (in \wn) of select diatomic molecules. The EOM-CCSD(SOC) results are taken from ref~\citenum{cao.2017.10.1039/C6CP07588F}; the SO-EOM-CCSD results are from ref~\citenum{cheng.2018.10.1063/1.5012041}; the BP1-DSRG-MRPT2 results are from ref~\citenum{wang.2025.10.3390/molecules30092082}; the BP1- and DKH2-NEVPT2 results are from ref~\citenum{majumder.2024.10.1021/acs.jctc.4c00458}; the DKH1-SO-L-PDFT results are from ref~\citenum{jangid.2025.10.1021/acs.jctc.5c01633}.}
\label{tab:diatomics}
\begin{threeparttable}
\begin{tabular}{lllllllBBll}
\hhline{===========}
Molecule 
& EOM-CCSD 
& SO-EOM-
& BP1-DSRG-
& BP1-
& DKH1-SO-
& DKH2-
& X2C-
& X2C-DSRG-
& Exp.
& ref \\
($r_{\mathrm{e}}$ / \AA)
& (SOC)\tnote{b,c}
& CCSD\tnote{d}
& MRPT2\tnote{a}
& NEVPT2
& L-PDFT
& NEVPT2
& CASSCF
& MRPT2\tnote{a}
&
& \\\hline

CH (1.1199)
& 27.40
& --
& 28.50\tnote{e}
& 29.00\tnote{e}
& --
& 27.30\tnote{e}
& 28.14
& 28.62
& 27
& \citenum{huber.1979.10.1007/978-1-4757-0961-2} \\

OH (0.96966)
& 140.10
& 136.30
& 149.20
& 152.50
& 138.00
& 123.20
& 137.80
& 134.42
& 139
& \citenum{huber.1979.10.1007/978-1-4757-0961-2} \\

SiH (1.5201)
& 139.30
& --
& 131.90\tnote{e}
& 128.00\tnote{e}
& --
& 135.60\tnote{e}
& 131.64
& 147.18
& 142
& \citenum{huber.1979.10.1007/978-1-4757-0961-2} \\

SH (1.3409)
& 375.30
& 373.80
& 374.40
& 375.60
& 348.00
& 378.20
& 381.43
& 386.68
& 377
& \citenum{huber.1979.10.1007/978-1-4757-0961-2} \\

GeH (1.588)
& 882.90
& --
& 864.10\tnote{e}
& 864.10\tnote{e}
& --
& 854.90\tnote{e}
& 802.40
& 913.82
& 892
& \citenum{huber.1979.10.1007/978-1-4757-0961-2} \\

SeH (1.464)
& 1742.90
& 1716.80
& 1832.70
& 1836.70
& 1637.00
& 1793.20
& 1765.87
& 1776.60
& 1763
& \citenum{ram.2000.10.1006/jmsp.2000.8147} \\

SnH (1.7815)
& 2187.00
& --
& 2311.80\tnote{e}
& 2286.30\tnote{e}
& --
& 2103.70\tnote{e}
& 1946.31
& 2276.55
& 2178
& \citenum{huber.1979.10.1007/978-1-4757-0961-2} \\

TeH (1.656)
& 3913.40
& 3751.70
& 4271.00
& 4293.50
& 3554.00
& 3956.50
& 3835.84
& 3966.43
& 3816
& \citenum{fink.1989.10.1016/0022-28528990094-5} \\

FO (1.354)
& --
& 193.60
& 180.00
& 180.00
& 192.00
& 189.20
& 199.05
& 200.53
& 196.6
& \citenum{miller.2001.10.1006/jmsp.2000.8257} \\

ClO (1.569)
& --
& 318.70
& 280.30
& 299.70
& 302.00
& 324.40
& 315.11
& 343.73
& 322
& \citenum{drouin.2001.10.1006/jmsp.2001.8332} \\

BrO (1.717)
& --
& 984.20
& 853.20
& 961.90
& 897.00
& 1012.00
& 784.86
& 988.10
& 975.4
& \citenum{drouin.2001.10.1006/jmsp.2000.8252} \\

IO (1.8676)
& --
& 2143.60
& 1871.70
& 2303.80
& 1950.00
& 2237.50
& 1458.23
& 2054.67
& 2091
& \citenum{gilles.1991.10.1063/1.461746} \\\hline

MAE
& 17.7\tnote{f}
& 23.0\tnote{f}
& 92.6
& 82.0
& 82.8\tnote{f}
& 41.6
& 99.5
& 31.7
& --
& \\

MAPE
& 1.2\%\tnote{f}
& 1.6\%\tnote{f}
& 7.5\%
& 6.6\%
& 5.7\%\tnote{f}
& 3.8\%
& 7.3\%
& 3.2\%
& --
& \\
\hhline{===========}
\end{tabular}
\begin{tablenotes}
\scriptsize
\item [a] A flow parameter of $s=0.5$ \sunit~was used for all DSRG computations.
\item [b] The sf-X2C+so-DKH1 scheme and the decontracted ANO-RCC-VTZP basis set were used for the EOM-CCSD(SOC) computations.
\item [c] The EOM-CCSD(SOC) computations for chalcogen hydrides (OH, SH, SeH, TeH) used bond lengths of 0.9697, 1.3409, 1.475, and 1.741 \AA, respectively, taken from ref~\citenum{huber.1979.10.1007/978-1-4757-0961-2}.
\item [d] The T-relaxed scheme was used for the SO-EOM-CCSD computations.
\item [e] The decontracted ANO-RCC-VTZP basis set was used for these computations.
\item [f] Unavailable data points have been omitted from averaging.
\end{tablenotes}
\end{threeparttable}
\end{table*}

We report in \cref{tab:diatomics} the spin--orbit splittings of various diatomic molecules, and we can indeed observe that the two single-reference methods do perform quite well for the molecules for which data is available.
Both the EOM-CCSD(SOC) and SO-EOM-CCSD methods are based on IP-EOM-CCSD and naturally capture the open-shell character of these molecules. The inclusion of SOC in the EOM step allows them to capture the mixing of the low-lying states by SOC.
Among the multireference methods, the X2C-DSRG-MRPT2 method performs the best, with a MAE of 31.7 \wn~and a MAPE of 3.2\%, which are comparable to the accuracy of the single-reference methods.

\subsection{Potential Energy Curves of Thallium Hydride}
\label{sec:pes}
\begin{figure}[!hbtp]
    \centering
    \includegraphics[width=3.25in]{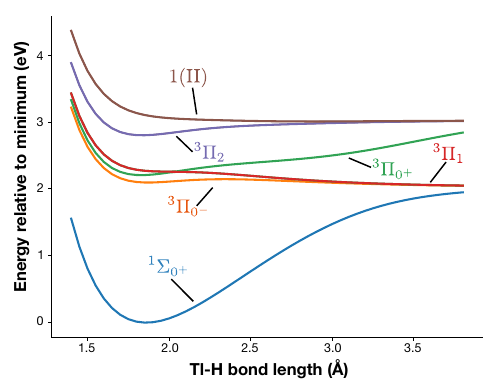}
    \caption{Potential energy curves of the TlH molecule computed with the X2C-DSRG-MRPT2 method and a decontracted ANO-RCC-VTZP basis set. The energy is relative to the minimum of the $^1\Sigma_{0^+}$ state.}
    \label{fig:tlh}
\end{figure}

The TlH molecule is a prototypical system where the interplay between strong SOC and electron correlation is expected to be significant, and it has been the subject of several previous computational studies.\cite{zeng.2010.10.1063/1.3297887,shiozaki.2015.10.1021/acs.jctc.5b00754,liu.2021.10.1016/j.jqsrt.2021.107667,lu.2022.10.1021/acs.jctc.2c00171}
The correct description of the potential energy surfaces of the low-lying electronic states across a wide range of bond lengths is furthermore a stringent test for the accuracy of multireference methods.
Here, we target the lowest six electronic states of the TlH molecule, namely, the $^1\Sigma_{0^+}$, $^3\Pi_{0^-}$, $^3\Pi_{0^+}$, $^3\Pi_1$, $^3\Pi_2$, and $1$(II) states, which have degeneracies of 1, 1, 1, 2, 2, and 2, respectively.
We have followed the state assignments of ref~\citenum{liu.2021.10.1016/j.jqsrt.2021.107667}.
The first two states and the $^3\Pi_1$ state dissociate to the Tl($^2$P$_{1/2}$) + H($^2$S$_{1/2}$) limit, while the remaining states dissociate to the Tl($^2$P$_{3/2}$) + H($^2$S$_{1/2}$) limit.
The latter atomic limit also contains the $1$(III) and $0^-$(II) states with degeneracies of 2 and 1, respectively, but they are not included in the computations here, as we are primarily interested in showing the more complex behavior of the lower-energy manifold of states.
The X2C-DSRG-MRPT2 computations were performed with the decontracted ANO-RCC-VTZP basis set,\cite{widmark.1990.10.1007/BF01120130,roos.2004.10.1021/jp031064+} while an active space of 4 electrons in 10 spinors was used, corresponding to the thallium $6s$ and $6p$ orbitals and the hydrogen $1s$ orbital, and the lowest 9 roots were included in the state-averaging procedure.
The flow parameter was set to $s=0.5$ \sunit, and 54 core spinors were frozen in the DSRG-MRPT2 step.
The Tl--H bond length was varied from 1.4 to 3.8 \AA~with a step size of 0.05 \AA.
We present the potential energy curves of the TlH molecule in \cref{fig:tlh}, which shows very good agreement with the accurate ic-MRCI+Q-SO results of ref~\citenum{liu.2021.10.1016/j.jqsrt.2021.107667} across the entire range of bond lengths.
The X2C-DSRG-MRPT2 curves also correctly capture the relative ordering of the states across the dissociation process.
Specifically, X2C-DSRG-MRPT2 correctly predicts that the $^3\Pi_{0^+}$ state is weakly bound, and that a shoulder feature exists in the PES of the $^3\Pi_{0^-}$ state, and that a crossing between the $^3\Pi_{0^+}$ and the $^3\Pi_1$ states takes place around 2.0 \AA.

For the computation of the spectroscopic constants shown in \cref{tab:tlh}, separate state-specific computations were performed for the $^1\Sigma_{0^+}$ state near the equilibrium geometry with all other parameters being the same.
The Tl--H bond length was varied from 1.78 to 1.96 \AA~with a step size of 0.005 \AA, and the spectroscopic constants $R_{\mathrm{e}}$, $\omega_{\mathrm{e}}$, and $E(r_{\mathrm{e}})$ were obtained by using the improved local interpolating moving least squares method (L-IMLS-G2) procedure as implemented in the \textsc{Psi4} package.\cite{smith.2020.10.1063/5.0006002}
A single point computation was performed at 10.00 \AA~to obtain the dissociation energy $D_{\mathrm{e}}$.
\begin{table}[!htbp]
\centering
\caption{Spectroscopic constants of the $^1\Sigma_{0^+}$ state of the $^{205}$TlH molecule computed with various methods. The X2C-CASSCF (Boettger) and uncontracted X2C-MRPT2 results are taken from ref~\citenum{lu.2022.10.1021/acs.jctc.2c00171}. The 4C-CASPT2 and 4C-ic-MRCI+Q results are from ref~\citenum{shiozaki.2015.10.1021/acs.jctc.5b00754}. The SO-MCQDPT results are from ref~\citenum{zeng.2010.10.1063/1.3297887}, and the ic-MRCI+Q-SO results are from ref~\citenum{liu.2021.10.1016/j.jqsrt.2021.107667}. The experimental results are taken from Refs.~\citenum{urban.1989.10.1016/0009-26148987368-3,grundstrom.1937.10.1038/140365b0,neuhaus.1959.10.1007/BF01337608,titov.2001.10.1002/1097-461X200181:6<409::AID-QUA1010>3.0.CO;2-0}.}
\label{tab:tlh}
\begin{tabular}{llll}
\hhline{====}
Method & $R_{\mathrm{e}}$ / \AA & $\omega_{\mathrm{e}}$ / cm$^{-1}$ & $D_{\mathrm{e}}$ / eV \\ \hline
X2C-CASSCF (Boettger) &1.924&1439.6&1.52 \\
\textbf{X2C-CASSCF (SNSO)} &\textbf{1.919}&\textbf{1295.6}& \textbf{1.21}\\
uc-X2C-MRPT2 &1.873&1395.4&2.06 \\
\textbf{X2C-DSRG-MRPT2} &\textbf{1.8706}&\textbf{1379.1}&\textbf{2.04} \\
4C-CASPT2 &1.870&--&1.84 \\
4C-ic-MRCI+Q & 1.872 & --&2.00 \\
SO-MCQDPT &1.876&1391&2.03 \\
ic-MRCI+Q-SO &1.8826&1390.4&2.05 \\\hline
Exp. &1.872&1390.7&2.06 \\
\hhline{====}
\end{tabular}
\end{table}
The computed spectroscopic constants are in very good agreement with the experimental values, and the X2C-DSRG-MRPT2 results are comparable to the more expensive uncontracted X2C-MRPT2 results of Li and co-workers.\cite{lu.2022.10.1021/acs.jctc.2c00171} 
Compared to the four-component methods, the X2C-DSRG-MRPT2 results are in better agreement with experimental values than the 4C-CASPT2 results of Shiozaki and co-workers,\cite{shiozaki.2015.10.1021/acs.jctc.5b00754} and are in line with the 4C-ic-MRCI+Q results, both of which used the four-component Dirac--Coulomb--Breit Hamiltonian, and employed state-specific computations with a basis set of similar quality and size and the same active space.
The perturbative SOC methods (SO-MCQDPT and ic-MRCI+Q-SO) perform quite well, but both used very large active spaces and bases, as well as relativistic effective core potentials, which can introduce systematic errors that are difficult to quantify.
Also notable is the significant improvement in the spectroscopic constants upon inclusion of dynamical correlation with the X2C-DSRG-MRPT2 method, compared to the X2C-CASSCF (SNSO) results, which highlights the importance of dynamical correlation for achieving accurate predictions of spectroscopic constants for systems containing heavy elements.

Before concluding this section, we briefly report the timing breakdown of a representative X2C-DSRG-MRPT2 computation for the TlH molecule, at $1.9$ \AA, in \cref{tab:timing}.
\begin{table}[!htbp]
\centering
\caption{Timing breakdown of an X2C-DSRG-MRPT2 computation for the TlH molecule at a bond length of 1.9 \AA.}
\label{tab:timing}
\begin{tabular}{ll}
\hhline{==}
Step & Time (s) \\ \hline
X2C-HF & 66.64 \\
X2C-CASSCF & 533.40 \\
X2C-DSRG-MRPT2 & 30.27 \\ \hline
Total & 630.31 \\
\hhline{==}
\end{tabular}
\end{table}
This computation was performed on a MacBook Air with an Apple M2 chip, with 4 performance cores and 4 efficiency cores, and 16 GB of RAM.
The RAM usage was around 1.2 GB, and consisted almost entirely of the three-index density-fitted integrals.
We can see that systematic and significant improvements in the accuracy of the computed spin--orbit splittings can be achieved using the X2C-DSRG-MRPT2 method without a significant increase in computational cost relative to the X2C-CASSCF method.
Therefore, we can conclude that the X2C-DSRG-MRPT2 method, with $\mathcal{O}(N^4)$ scaling and the density fitting approximation, is both accurate and highly efficient in terms of computational cost and memory usage.

\section{Conclusion}
\label{sec:conclusion}
In this work, we introduced the X2C-DSRG-MRPT2 method for the efficient and accurate inclusion of SOC in multireference perturbation theory.
A comparison of different approximation schemes highlights the importance of performing orbital optimization in the presence of SOC to achieve systematic improvements in the accuracy of the X2C-DSRG-MRPT2 computed spin--orbit splittings across the entire periodic table.
We have applied X2C-DSRG-MRPT2 to the computation of spin--orbit splittings of a broad set of atomic and molecular systems containing up to sixth-row elements,  comparing results with experimental data and other computational methods.
Our results show that the X2C-DSRG-MRPT2 method is competitive with the most accurate multireference approaches available for computing spin--orbit splittings.
The X2C-DSRG-MRPT2 method, implemented here efficiently with the density fitting approximation, incurs only a modest increase in computational cost relative to the underlying X2C-CASSCF reference wave function, and therefore it is a promising method for the accurate and efficient inclusion of SOC effects in strongly correlated systems containing elements across the periodic table.
Future work will focus on the extension of X2C-DSRG-MRPT2 to the third-order perturbation theory (DSRG-MRPT3),\cite{li.2017.10.1063/1.4979016} which has been shown to provide significant and consistent improvements in accuracy over DSRG-MRPT2 in the four-component case.\cite{zhao.2024.10.1021/acs.jpclett.4c01372}
Work in these directions is currently underway in our group.

\section*{Data Availability Statement}
The X2C-CASSCF and X2C-DSRG-MRPT2 methods are implemented in the \textsc{Forte2} program package,\cite{forte2} which is freely available at \url{https://github.com/evangelistalab/forte2}.
All data generated in this study are available in the supplementary material.

\section*{Acknowledgments}
This research was supported by the U.S. Department of Energy under Award DE-SC0024532. 
We acknowledge the use of ChatGPT Pro (OpenAI) to improve the grammar, phrasing, and formatting of the manuscript.

\section*{Supplementary Material}
The supplementary material contains: 1) all the zero-field splittings of the atomic and diatomic species considered in this work computed at the X2C-CASSCF and X2C-DSRG-MRPT2 and approximate variants thereof, and 2) potential energy curves of the TlH molecule computed with state-averaged and state-specific X2C-DSRG-MRPT2.

\section*{Author declarations}
\subsection*{Conflict of interest}
The authors have no conflicts to disclose.
\subsection*{Author contributions}
\textbf{Zijun Zhao}: Conceptualization (lead), Methodology (lead), Software (lead), Validation (lead), Formal analysis (lead), Investigation (lead), Data curation (lead), Writing -- original draft (lead), Writing -- review \& editing (equal), Visualization (lead).
\textbf{Francesco A. Evangelista}: Conceptualization (equal), Methodology (equal), Software (equal), Validation (equal), Formal analysis (equal), Investigation (equal), Data curation (equal), Writing -- original draft (equal), Writing -- review \& editing (equal), Visualization (equal).

\appendix
\setcounter{equation}{0}
\renewcommand{\theequation}{A\arabic{equation}}
\section{The DSRG-MRPT2 Formalism}
\label{sec:dsrg}
The multireference driven similarity renormalization group (MR-DSRG) is a robust and systematically improvable formalism for the treatment of strongly correlated molecular systems at low polynomial scaling.\cite{li.2019.10.1146/annurev-physchem-042018-052416}
The detailed derivations and analyses of the DSRG-MRPT2 method\cite{evangelista.2014.10.1063/1.4890660,li.2015.10.1021/acs.jctc.5b00134} and its state-averaged extension for the treatment of valence excited states\cite{li.2018.10.1063/1.5019793} have been presented in several previous works; therefore, we will only summarize the key results.
The MR-DSRG uses a complete active space (CAS) reference wave function [\cref{eq:cas_ref}].
The \emph{model space} is defined as the span of all determinants $\mathcal{M}=\{\ket{\Phi_{\mu}}\}$, $\mu=1,\ldots,d$ in the CAS.
In the following, we will use indices $m,n,\dots$ to denote core ($\bvec{C}$) orbitals, $u,v,\dots$ for active ($\bvec{A}$) orbitals, $e,f,\dots$ for virtual ($\bvec{V}$) orbitals, and $i,j,\dots$ for hole ($\bvec{H}=\bvec{C}\cup\bvec{A}$) orbitals, $a,b,\dots$ for particle ($\bvec{P}=\bvec{A}\cup\bvec{V}$) orbitals, and $p,q,r,s,\dots$ for general ($\bvec{G}$) orbitals.
The second-quantized Hamiltonian in the normal ordered form with respect to the reference $\ket{\Psi_0}$ is given by
\begin{equation}
    \hat{H} = E_0 + \sum_{pq}^{\bvec{G}} f_p^q \{\hat{a}_q^p\} + \frac{1}{4} \sum_{pqrs}^{\bvec{G}} v_{pq}^{rs} \{\hat{a}_{rs}^{pq}\} \equiv E_0 + \hat{F} + \hat{V},
\end{equation}
where the operator $\hat{a}^{ab\dots}_{ij\dots} = \hat{a}\adj_a \hat{a}\adj_b \cdots \hat{a}_j \hat{a}_i$ is a string of second-quantized creation and annihilation operators, while the curly braces $\{\cdots\}$ denote generalized normal ordering (GNO) with respect to the multi-determinantal reference $\ket{\Psi_0}$,\cite{kutzelnigg.1997.10.1063/1.474405,mukherjee.1997.10.1016/S0009-26149700714-8} $E_0 = \braket{\Psi_0|\hat{H}|\Psi_0}$ is the reference energy, $v_{pq}^{rs} = \braket{rs||pq}$ are the antisymmetrized two-electron integrals, and $f_p^q$ are the matrix elements of the generalized Fock operator defined as
\begin{equation}
    f_p^q = h_p^q + \sum_{rs}^{\bvec{G}} v_{pr}^{qs} \gamma_s^r,
\end{equation}
where $h_p^q=\braket{q|\hat{h}|p}$ are the one-electron integrals, and $\gamma^r_s = \braket{\Psi_0|\hat{a}^r_s|\Psi_0}$ are the one-body reduced density matrix (1-RDM) elements.

The MR-DSRG is formulated in terms of a continuous unitary transformation, controlled by a time-like \emph{flow parameter}, $s\in[0,\infty)$, described by
\begin{equation}
    \label{eq:dsrg_trans}
    \bar{H}(s) = e^{-\hat{A}(s)} \hat{H} e^{\hat{A}(s)},
\end{equation}
where $\bar{H}(s)$ is the similarity-transformed Hamiltonian, and $\hat{A}(s)$ is an anti-Hermitian operator defined as $\hat{A}(s)=\hat{T}(s)-\hat{T}\adj(s)$, with $\hat{T}(s)$ being a cluster operator given by $\hat{T}(s)=\sum_{k=1}^{N} \hat{T}_k(s)$, where $N$ is the maximum excitation rank, and $\hat{T}_k(s)$ is the $k$-body excitation operator defined as
\begin{equation}
    \hat{T}_k(s) = \frac{1}{(k!)^2} \sum_{ij\dots}^{\bvec{H}} \sum_{ab\dots}^{\bvec{P}} t^{ij\dots}_{ab\dots}(s) \{\hat{a}^{ab\dots}_{ij\dots}\}.
\end{equation}
The amplitudes $t^{ij\dots}_{ab\dots}(s)$ are determined such that the unitary transformation [\cref{eq:dsrg_trans}] decouples the model space from the rest of the Hilbert space in the limit $s\rightarrow\infty$.
Since the off-diagonal many-body elements of $\bar{H}(s)$, $\bar{H}^{\mathrm{N}}(s)$, are responsible for this coupling, the MR-DSRG amplitudes are obtained by the following set of \emph{many-body conditions}
\begin{equation}
    \label{eq:mb_conds}
    \bar{H}^{\mathrm{N}}(s) = \hat{R}(s),
\end{equation}
where $\hat{R}(s)$ is a \emph{source operator} that drives the renormalization process such that $\hat{R}(0)=\bar{H}^{\mathrm{N}}(0)$, and $\lim_{s\rightarrow\infty}\hat{R}(s)=\hat{0}$.
An analytical form of the source operator has been derived in previous works.\cite{evangelista.2014.10.1063/1.4890660,li.2015.10.1021/acs.jctc.5b00134}

The DSRG-MRPT2 method is obtained by using the M{\o}ller--Plesset partitioning of the Hamiltonian, where the zeroth-order Hamiltonian is defined as
\begin{equation}
    \hat{H}^{(0)} = E_0 + \sum_{p}^{\bvec{G}} \epsilon_p \{\hat{a}_p^p\},
\end{equation}
with $\epsilon_p = f_p^p$ being the diagonal elements of the generalized Fock matrix \emph{in the semi-canonical basis}, while the perturbation is given by $\hat{H}^{(1)} = \hat{F}^{(1)}+\hat{V}^{(1)} \equiv \hat{H} - \hat{H}^{(0)}$, where $\hat{F}^{(1)}=\hat{F}-\hat{H}^{(0)}$ and $\hat{V}^{(1)}=\hat{V}$.
The first non-vanishing contribution to the energy comes in at the second order in perturbation theory, given by (the flow parameter $s$ is omitted henceforth for clarity)
\begin{align}
    \label{eq:dsrg_mrpt2_energy}
    E^{(2)} = &\braket{\Psi_0|[\hat{H}^{(1)}, \hat{A}^{(1)}]|\Psi_0} \\
    &+ \frac{1}{2}\braket{\Psi_0|[[\hat{H}^{(0)}, \hat{A}^{(1)}], \hat{A}^{(1)}]|\Psi_0}.
\end{align}
The above expression can be formulated purely in terms of contractions among rank-2 and -4 tensors containing dressed Hamiltonian integrals and amplitudes, and up to three-body reduced density cumulants.
This can be done by defining an appropriate first-order dressed Hamiltonian, $\tilde{H}^{(1)} = [\hat{H}^{(0)}, \hat{A}^{(1)}] + 2\hat{H}^{(1)}$, whose one- and two-body elements are given by
\begin{align}
\begin{split}
    &\tilde{f}_a^{i,(1)} = f_a^{i,(1)}[1+e^{-s(\Delta_a^i)^2}] + \sum_{ux}^{\bvec{A}} \Delta_u^x t_{ax}^{iu,(1)} \gamma_u^x e^{-s(\Delta_a^i)^2},\\
    &\tilde{v}_{ab}^{ij,(1)} = v_{ab}^{ij,(1)}[1+e^{-s(\Delta_{ab}^{ij})^2}],
\end{split}
\end{align}
where $\Delta_{ab\dots}^{ij\dots} = \epsilon_i + \epsilon_j + \cdots - \epsilon_a - \epsilon_b - \cdots$ are the generalized M{\o}ller--Plesset energy denominators.
The complex-conjugation symmetry $\tilde{f}_i^{a,(1)} = (\tilde{f}_a^{i,(1)})^*$ and $\tilde{v}_{ij}^{ab,(1)} = (\tilde{v}_{ab}^{ij,(1)})^*$ can be used to obtain the remaining elements of $\tilde{H}^{(1)}$.
With the above definition of $\tilde{H}^{(1)}$, the DSRG-MRPT2 energy can be expressed compactly as 
\begin{equation}
    \label{eq:pt2_commutator}
    E^{(2)} = \frac{1}{2}\braket{\Psi_0|[\tilde{H}^{(1)}, \hat{A}^{(1)}]|\Psi_0},
\end{equation}
where the first-order perturbative amplitudes are given by
\begin{align}
    &t^{ij,(1)}_{ab} = v_{ab}^{ij,(1)} \frac{1-e^{-s(\Delta_{ab}^{ij})^2}}{\Delta_{ab}^{ij}},\\
    &t^{i,(1)}_{a} = \left(f^{i,(1)}_{a}+\sum_{ux}^{\bvec{A}}\Delta^x_ut^{iu,(1)}_{ax}\gamma^x_u \right)\frac{1-e^{-s(\Delta_{a}^{i})^2}}{\Delta_{a}^{i}}.
\end{align}

The expectation value of the commutator in \eqref{eq:pt2_commutator} can be evaluated using Wick's theorem for a multi-determinantal reference, and the resulting tensor contractions are given by, in the spinor basis:
\begin{align}
\begin{split}
E^{(2)} = & t^{m,(1)}_{v} \tilde{f}_{m}^{u,(1)} \eta^v_u
    + t^{m,(1)}_{a} \tilde{f}_{m}^{a,(1)}
    + t^{v,(1)}_{e} \tilde{f}_{u}^{e,(1)} \gamma^u_v\\
    &-\frac{1}{2} t^{mx,(1)}_{vw} \tilde{f}_{m}^{u,(1)} \lambda^{vw}_{ux}
    -\frac{1}{2} t^{wx,(1)}_{ve} \tilde{f}_{u}^{e,(1)} \lambda^{uv}_{wx}\\
    &-\frac{1}{2} t^{m,(1)}_{u} \tilde{v}_{mv}^{wx,(1)} \lambda^{uv}_{wx}
    -\frac{1}{2} t^{u,(1)}_{e} \tilde{v}_{vw}^{xe,(1)} \lambda^{vw}_{ux}\\
    &+\frac{1}{4} t^{mn,(1)}_{ef} \tilde{v}_{mn}^{ef,(1)}
    +\frac{1}{2} t^{mu,(1)}_{ef} \tilde{v}_{mv}^{ef,(1)} \gamma^v_u\\
    &+\frac{1}{2} t^{mn,(1)}_{ue} \tilde{v}_{mn}^{ve,(1)} \eta^u_v
    +\frac{1}{4} t^{mn,(1)}_{uv} \tilde{v}_{mn}^{wx,(1)} \eta^v_x \eta^u_w\\
    &+\frac{1}{8} t^{mn,(1)}_{uv} \tilde{v}_{mn}^{wx,(1)} \lambda^{uv}_{wx}
    +\frac{1}{2} t^{mw,(1)}_{uv} \tilde{v}_{mx}^{yz,(1)} \eta^v_z \eta^u_y \gamma^x_w\\
    &+ t^{mw,(1)}_{uv} \tilde{v}_{mx}^{yz,(1)} \eta^v_z \lambda^{ux}_{wy}
    +\frac{1}{4} t^{mw,(1)}_{uv} \tilde{v}_{mx}^{yz,(1)} \gamma^x_w \lambda^{uv}_{yz}\\
    &+\frac{1}{4} t^{mw,(1)}_{uv} \tilde{v}_{mx}^{yz,(1)} \lambda^{uvx}_{wyz}
    + t^{mv,(1)}_{ue} \tilde{v}_{mw}^{xe,(1)} \eta^u_x \gamma^w_v\\
    &+ t^{mv,(1)}_{ue} \tilde{v}_{mw}^{xe,(1)} \lambda^{uw}_{vx}
    +\frac{1}{2} t^{vw,(1)}_{ue} \tilde{v}_{xy}^{ze,(1)} \eta^u_z \gamma^y_w \gamma^x_v\\
    &+\frac{1}{4} t^{vw,(1)}_{ue} \tilde{v}_{xy}^{ze,(1)} \eta^u_z \lambda^{xy}_{vw}
    + t^{vw,(1)}_{ue} \tilde{v}_{xy}^{ze,(1)} \gamma^y_w \lambda^{ux}_{vz}\\
    &-\frac{1}{4} t^{vw,(1)}_{ue} \tilde{v}_{xy}^{ze,(1)} \lambda^{uxy}_{vwz}
    +\frac{1}{4} t^{uv,(1)}_{ef} \tilde{v}_{wx}^{ef,(1)} \gamma^x_v \gamma^w_u\\
    &+\frac{1}{8} t^{uv,(1)}_{ef} \tilde{v}_{wx}^{ef,(1)} \lambda^{wx}_{uv}.
\end{split}
\end{align}
In \cref{eq:dsrg_mrpt2_energy}, $\lambda^{uv}_{wx}=\gamma^{uv}_{wx}-\gamma^u_w \gamma^v_x + \gamma^u_x \gamma^v_w$ is the two-body reduced density cumulant, and the three-body reduced density cumulant $\lambda^{uvw}_{xyz}$ is defined as:
\begin{align}
\begin{split}
\lambda^{uvw}_{xyz} = &\gamma^{uvw}_{xyz} \\
 &-\gamma^u_x \gamma^{vw}_{yz}
 -\gamma^v_y \gamma^{uw}_{xz}
 -\gamma^w_z \gamma^{uv}_{xy}\\
 &+\gamma^u_y \gamma^{vw}_{xz}
 +\gamma^u_z \gamma^{vw}_{yx}
 +\gamma^v_x \gamma^{uw}_{yz}\\
 &+\gamma^v_z \gamma^{uw}_{xy}
 +\gamma^w_x \gamma^{uv}_{zy}
 +\gamma^w_y \gamma^{uv}_{xz}\\
 &+2( \gamma^u_x \gamma^v_y \gamma^w_z
 - \gamma^u_x \gamma^v_z \gamma^w_y
 - \gamma^u_z \gamma^v_y \gamma^w_x\\
 &- \gamma^u_y \gamma^v_x \gamma^w_z
 + \gamma^u_y \gamma^v_z \gamma^w_x
 + \gamma^u_z \gamma^v_x \gamma^w_y).
\end{split}
\end{align}
The term $\frac{1}{4} t^{mn,(1)}_{ef} \tilde{v}_{mn}^{ef,(1)}$ is responsible for the $\mathcal{O}(N_{\mathrm{C}}^2 N_{\mathrm{V}}^2)$ asymptotic scaling in the common case of $N_{\mathrm{A}} \ll N_{\mathrm{C}} < N_{\mathrm{V}}$, and the term $-\frac{1}{4} t^{vw,(1)}_{ue} \tilde{v}_{xy}^{ze,(1)} \lambda^{uxy}_{vwz}$ is responsible for the $\mathcal{O}(N_{\mathrm{A}}^6N_{\mathrm{V}})$ scaling with respect to the number of active space spinors.
In the density-fitted implementation of the X2C-DSRG-MRPT2 method, the terms $\frac{1}{4} t^{mn,(1)}_{ef} \tilde{v}_{mn}^{ef,(1)}$, $\frac{1}{2} t^{mu,(1)}_{ef} \tilde{v}_{mv}^{ef,(1)} \gamma^v_u$, and $\frac{1}{2} t^{mn,(1)}_{ue} \tilde{v}_{mn}^{ve,(1)} \eta^u_v$ are evaluated on-the-fly using the three-index density-fitted integrals to avoid the explicit formation and storage of the four-index two-electron integral and first-order amplitude blocks, which can become memory bottlenecks for large systems.\cite{hannon.2016.10.1063/1.4951684}
The remaining terms are evaluated using the standard four-index two-electron integral blocks, which are computed and stored in memory.

The fact that the energy can be evaluated entirely in terms of tensor contractions makes it possible to use optimized numerical libraries such as \texttt{numpy.einsum}\cite{harris.2020.10.1038/s41586-020-2649-2} or \texttt{opt\_einsum},\cite{g.a.smith.2018.10.21105/joss.00753} making the method computationally affordable for moderate active spaces.

The computation of valence excited states can be achieved \emph{via} the state-averaging formalism,\cite{li.2018.10.1063/1.5019793} where the normal ordering is instead defined with respect to an ensemble of CASCI states of interest, $\mathbb{E}=\{\ket{\Psi_k},k=1,2,\dots,n\}$
The only changes required are the use of a state-averaged CASSCF (SA-CASSCF) [\cref{eq:sa_energy}] reference wave function; and the use of ensemble-averaged reduced density matrices ($\bvec{\bar{\gamma}}$) in the evaluation of the generalized Fock matrix and the MR-DSRG energy expressions, given by
\begin{equation}
    \bar{\gamma}_m = \sum_{k=1}^{n} w_k \gamma_m^{(k)},
\end{equation}
where $\gamma_m^{(k)}$ is the $m$-body RDM of state $k$.
To obtain the correlated states, the DSRG-MRPT2 similarity-transformed Hamiltonian [\cref{eq:dsrg_trans}] is diagonalized in the model space.
This has the additional benefit of accounting for reference relaxation in the presence of dynamic correlation, and can be done iteratively if desired.
The active-space $\bar{H}$ can be obtained cheaply alongside the energy evaluation routine [see Appendix B of ref~\citenum{li.2017.10.1063/1.4979016}].

\section{The Role of Dynamical Correlation}
\label{sec:dynamical_correlation}
\begin{figure}[!htbp]
    \centering
    \includegraphics[width=3.25in]{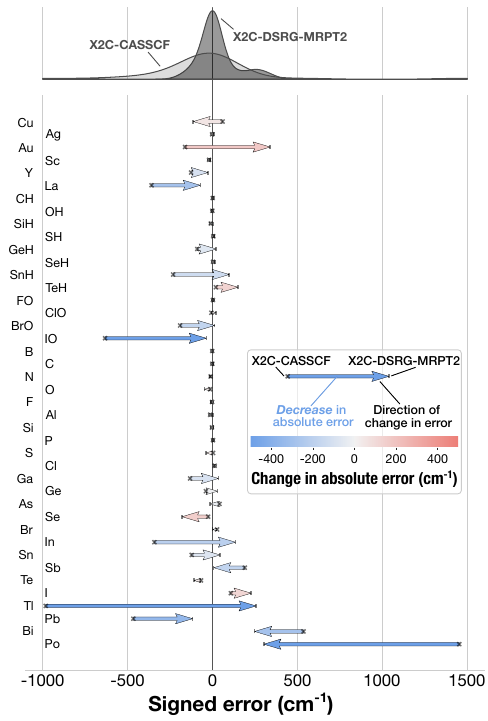}
    \caption{Comparison of the errors (in \wn) of X2C-CASSCF and X2C-DSRG-MRPT2 for the spin--orbit splittings of all systems considered in this work. The arrows point from the X2C-CASSCF results to the X2C-DSRG-MRPT2 results, and the color of the arrows indicates the change in the absolute error relative to experimental values. The top panel shows the error distributions for the two methods as kernel density estimates.}
    \label{fig:quiver}
\end{figure}

In this section, we analyze the role of dynamical correlation in the accurate prediction of spin--orbit splittings.
In \cref{fig:quiver}, we compare the errors of the X2C-CASSCF and X2C-DSRG-MRPT2 methods for the spin--orbit splittings of all systems considered so far, where the arrows point from the X2C-CASSCF results to the X2C-DSRG-MRPT2 results, and the color of the arrows indicates the change in the absolute error relative to experiment.
From the figure, it is clear that the inclusion of dynamical correlation through the X2C-DSRG-MRPT2 method significantly improves the accuracy of the computed spin--orbit splittings for most systems, and often the effect is quite large for systems containing heavier elements, for example, IO, Tl, and Po.
At the top of the figure, we can see that, although both X2C-CASSCF and X2C-DSRG-MRPT2 give essentially unbiased error distributions for the splittings, the X2C-DSRG-MRPT2 results are significantly more tightly clustered around zero error, while the X2C-CASSCF results show a much wider spread of errors.
This result shows the systematic improvement in the accuracy of the computed spin--orbit splittings upon inclusion of dynamical correlation with the X2C-DSRG-MRPT2 method, and highlights the importance of dynamical correlation in achieving accurate predictions of spin--orbit splittings, especially for systems containing heavier elements, where the interplay between strong SOC and dynamical correlation becomes more significant.

Tangentially, the effect of dynamical correlation for the approximate schemes presented in \cref{sec:pblock} is more mixed, as is apparent by comparing the errors of the sf-X2C-CASSCF-SO scheme with the sf-X2C-DSRG-MRPT2-SO (C) scheme.
As discussed, the latter scheme can be viewed as adding inconsistent perturbative corrections, which can lead to a deterioration in the accuracy of the computed splittings relative to the former scheme.

\bibliography{main}
\end{document}